\documentclass[useAMS,usenatbib]{mn2e}
\usepackage{times,amsmath,amssymb,graphicx,natbib,lscape}
\usepackage{longtable,supertabular,multirow}
\usepackage[figuresright]{rotating}

\newcommand{\cmmthree}{\mbox{cm$^{-3}$}}
\newcommand{\cmmtwo}{\mbox{cm$^{-2}$}}

\newcommand{\kms}{\mbox{km\,s$^{-1}$}}
\newcommand{\degree}{\mbox{$^{\circ}$}}
\newcommand{\msun}{\mbox{M$_{\odot}$}}

\newcommand{\microns}{\mbox{$\mu$m}}

\newcommand{\chthreecn}{\mbox{CH$_3$CN}}
\newcommand{\chthreeoh}{\mbox{CH$_{3}$OH}}
\newcommand{\hcop}{\mbox{HCO$^+$}}
\newcommand{\hthirteencop}{\mbox{H$^{13}$CO$^+$}}

\newcommand{\ntwohp}{\mbox{N$_2$H$^+$}}

\newcommand{\thirteenco}{\mbox{$^{13}$CO}}

\newcommand{\rarr}{\rightarrow}

\newcommand{\uchii}{UCH{\scriptsize II}}
\newcommand{\hii}{H{\scriptsize II}}

\title[Physical and chemical conditions in methanol maser selected
  hot-cores and UCH{\Large II} regions]{Physical and chemical
  conditions in methanol maser selected hot-cores and UCH{\Large II}
  regions } 
\author[C. R. Purcell {\it et al.}]{C. R. Purcell$^{1,2}$,
  S. N. Longmore$^{2,3,4}$
  M. G. Burton$^{2}$, 
  A. J. Walsh$^{2,5}$,
  V. Minier$^{2,6,7}$,
  \newauthor 
  M. R. Cunningham$^{2}$,
  R. Balasubramanyam$^{2,8}$ \\
$^{1}$~University of Manchester, Jodrell Bank Observatory,
  Macclesfield, Cheshire SK11 9DL, UK.\\
$^{2}$~School of Physics, University of New South Wales, Sydney, NSW 2052,
  Australia\\
$^{3}$~Harvard-Smithsonian Centre For Astrophysics, 60 Garden Street,
  Cambridge, MA, 02138, USA\\
$^{4}$~CSIRO Australia Telescope National Facillity, PO Box 76, Epping,
NSW 1710, Australia\\
$^{5}$~Centre for Astronomy, James Cook University, Townsville, QLD
  4811, Australia\\ 
$^{6}$~Service d'Astrophysique, DAPNIA/DSM/CEA Saclay, 91191
  Gif-sur-Yvette, France\\
$^{7}$~AIM, Unit\'e Mixte de Recherche, CEA$-$CNRS$-$Universit\'e
  Paris VII, UMR 7158, CEA/Saclay, 91191 Gif sur Yvette, France\\
$^{8}$~Raman Research Institute, Sadashivanagar, Bangalore 560 080,
  India\\}

\begin{document}

\pagerange{\pageref{firstpage}--\pageref{lastpage}} \pubyear{2006}

\maketitle

\label{firstpage}

\begin{abstract}
We present the results of a targeted 3-mm spectral line survey towards
the eighty-three 6.67\,GHz methanol maser selected star forming clumps
observed by \citet{Purcell2006}. In addition to the previously reported
measurements of \hcop\,(1\,--\,0), \hthirteencop\,(1\,--\,0), and
\chthreecn\,(5\,--\,4) \& (6\,--\,5), we used the Mopra antenna to
detect emission lines of \ntwohp\,(1\,--\,0), HCN\,(1\,--\,0) and
HNC\,(1\,--\,0) towards 82/83 clumps (99 per cent), and
\chthreeoh\,(2\,--\,1) towards 78/83 clumps (94 per cent).

The molecular line data have been used to derive virial and LTE
masses, rotational temperatures and chemical abundances in the clumps,
and these properties have been compared between sub-samples associated
with different indicators of evolution. The greatest differences are
found between clumps associated with 8.6\,GHz radio emission,
indicating the presence of an Ultra-Compact \hii~region,
and `isolated' masers (without associated radio emission), and between 
clumps exhibiting \chthreecn~emission and those without. In
particular, thermal \chthreeoh~is found to be brighter and more
abundant in Ultra-Compact \hii~(\uchii) regions and in
sources with detected \chthreecn, and may constitute a crude molecular clock in single dish observations.

Clumps associated with 8.6\,GHz radio emission tend to be more
massive {\it and} more luminous than clumps without radio
emission. This is likely because the most massive clumps evolve so
rapidly that a Huper-Compact\hii~or \uchii~region is the first visible tracer of
star-formation.

The gas-mass to sub-mm/IR luminosity relation for the combined sample
was found to be  L$\propto$\,M$^{0.68}$, considerably shallower than expected
for massive main-sequence stars. This implies that the mass of the
clumps is comparable to, or greater than the mass of the stellar
content. 

We find also that the mass of the hot core is correlated with the mass
of the clump in which it is embedded.
\end{abstract}



\section{Introduction}
Molecular emission is a powerful tool when used to investigate the
physical and chemical conditions in hot cores. Transitions requiring
different temperatures and densities for excitation constitute an
excellent probe of physical structure. Because the chemical properties
of hot cores vary with time, the relative molecular abundances can
also be used as indicators of evolution. 

In the preceding decades, representative line surveys of a limited
sample of cores have begun to identify the molecules most suited to
investigating the process of massive star formation. To date these
have been restricted to a few objects: Orion-KL
(\citealt{Blake1986,Turner1989,Ziurys1993,Schilke1997}), Sgr-B2
(\citealt{Cummins1986,Turner1989,Sutton1991}), G34.3+0.15
(\citealt{MacDonald1996,Kim2000,Kim2001}), IRAS 17470$-$2853 (also
known as G5.89$-$0.39, \citealt{Kim2002}), G305.2+0.2, \citep{Walsh2006}.

Chemical models, including both gas-phase and
grain-surface reactions have been developed for specific objects whose
physical structure and conditions are well known, e.g. G34.3+0.15
\citep{Millar1997b,Thompson1999}, as well as for general cases,
e.g. \citet{Rodgers2003}. These and earlier models are limited by
the computational power available at the time and consider
either time-dependent chemistry  at a test position, or a `snapshot'
of spatial abundances at a single time. Advances in computational
power have recently allowed the development of models which consider
both time- {\it and}
space-dependent chemistry simultaneously, e.g., the model of
AFGL\,2591 by \citet{Doty2002}. Today it is the lack of observational
constraints which restrict our understanding of chemistry in young
massive stellar objects. The priority in the coming years must be to
assemble a large sample of molecular abundance measurements towards
a broad range of massive star forming regions, at different stages of
evolution and with luminosities and masses from B3 to O5 type massive
stars.

In this paper we present the remaining results of the Mopra `Hot
Molecular Cores' (HMC) survey towards 6.67\,GHz methanol (\chthreeoh) maser
selected targets. The initial results were presented in
\citet{Purcell2006} (hereafter Paper 1), wherein we
reported the detection of the hot-core species methyl-cyanide
(\chthreecn) and the observations of the formyl ion (\hcop), and its
isotopomer (\hthirteencop). Here we expand the analysis to cover
thermal transitions of methanol (\chthreeoh), carbon-monoxide
(\thirteenco), diazenylium (\ntwohp), hydrocyanic acid (HCN) and
hydroisocyanic acid (HNC). We analyse the full complement of molecules
for evidence of evolution. 

Note: Tables~3-9 are available in electronic form only along with
additional supporting text and figures.


\section{Observations}
Observations were conducted on the Mopra millimetre wave telescope
over five years 2000\,--\,2004 during the winter observing season,
June\,--\,September. We observed the transitions \chthreeoh\,(2\,--\,1),
\thirteenco\,(1\,--\,0), \ntwohp\,(1\,--\,0), HCN\,(1\,--\,0) 
and HNC\,(1\,--\,0) as single pointings, targeted at the maser
sites. Electronic properties of the molecules and transitions are
presented in Table~\ref{tab:transitions}. Five sources do not contain
methanol masers but were detected as discrete clumps at 1.2\,mm,
adjacent to clumps containing \chthreeoh~masers
\citep{Hill2005}. These `mm-only' sources were observed as potential
precursors to the hot-core phase of massive starformation. In each
case we targeted 
the peak of the 450\,\micron~continuum emission later detected by
\citet{Walsh2003}. The signal from the receiver was processed in an
autocorrelator backend configured to have a bandwidth of 64\,MHz,
split into 1024 channels. At 3-mm wavelengths this delivers a velocity
resolution of $\sim$0.2\,\kms~over a usable velocity interval of
$\sim$130\,\kms. 

All observations were performed in position switching mode by
integrating for equal times on the science object and an emission-free
reference position. Initial reduction of the raw data was performed
using the {\footnotesize SPC}\footnote{SPC is a spectral line
  reduction package for single dish telescopes, written by the
  ATNF. http://www.atnf.csiro.au/software/} package. Polynomial
baselines were subtracted from the line-free channels using the
XS\footnote{XS is written by P. Bergman, Onsala Space
  Observatory. ftp://yggdrasil.oso.chalmers.se/pub/xs/} 
software package. The {\footnotesize CLASS}\footnote{CLASS is part of the
  GILDAS (Grenoble Image and Line Data Analysis Software) working
  group software. http://www.iram.fr/IRAMFR/GILDAS/} spectral line
analysis software was used for all further processing and analysis,
including baseline subtraction and fitting Gaussian line profiles. For
further details on the observing setup and data analysis please refer
to Section 3 of Paper 1.


\section{Results}\label{sec:hmc_2results}
\begin{figure}
  \begin{center}
    \includegraphics[width=8.0cm, angle=0]{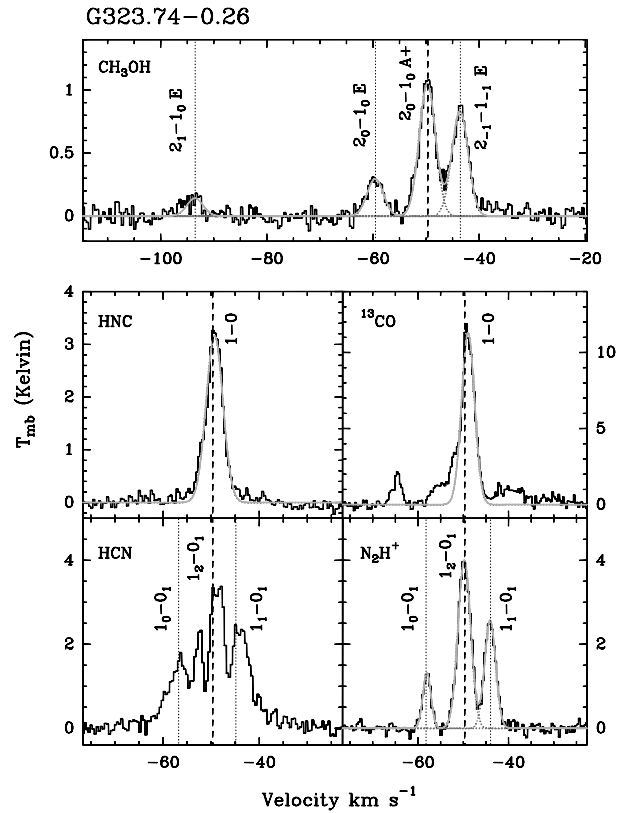}
    \caption[Sample spectra for the maser site associated with
    G323.74$-$0.26.]{\small~Sample spectra for the maser site associated with
    G323.74$-$0.26. The vertical dashed line marks the systemic velocity
    of the source, measured from the peak of the optically thin
    \hthirteencop~line profile (see
    \citealt{Purcell2006}). Vertical dotted lines mark the expected
    velocity of other transitions of the same molecule. Individual
    Gaussian fits to the line profiles are plotted by a dotted grey
    line and the cumulative fit by a solid grey line. We did not
    attempt to fit Gaussians to HCN lines as the majority of profiles
    have complex shapes.}
    \label{fig:spectra_example}
  \end{center}
\end{figure}
    Thermal molecular emission was detected in 82/83 targets,
with the exception being the source G10.10$-$0.72. In the interim
period this source has been reobserved at 6.67\,GHz using the 30\,m
Ceduna radio telescope (University of Tasmania) and no
\chthreeoh~maser was detected (S. Ellingsen, private communication). This
source is thus most likely a noise artifact in the \citet{Walsh1998}
survey and is discounted in the remainder of this work. Emission 
from \thirteenco(1\,--\,0), \ntwohp\,(1\,--\,0), HNC\,(1\,--\,0) and
HCN\,(1\,--\,0) was detected towards the remaining 82 sources, while 
\chthreeoh\,(2\,--\,1) emission was detected towards 79/82 sources (96 per
cent). The 3-$\sigma$ noise
limit on the spectra where no lines were detected is recorded in
Table~\ref{tab:nond_3sig}. Figure~\ref{fig:spectra_example} presents
sample  spectra for the source G323.74$-$0.26. The vertical
dashed line marks the systemic velocity of the gas, measured from
the optically thin \hthirteencop~line (Paper 1). In the case of 
\chthreeoh, the rest-frame is 
centred on the A+ transition, and for \ntwohp~and HCN the rest-frame
is centred on the ${\rm J_F}\,=1_2\rightarrow0_1$ E transition. Plots
of all molecular spectra for all 83 observed sources (including
  G10.10$-$0.72) are available as additional online material.


\subsection{Line profile parameters}\label{sec:line_profile_params}
As with \chthreecn~in Paper 1, we fit the
multiple transitions of \chthreeoh\,(2\,--\,1) simultaneously with
four Gaussian lines having equal full-width half-maximum (FWHM)
linewidths and whose line-centre separations were fixed to the
theoretical values. The weakest component in the spectrum, the ${\rm
  J_{K,K^{\prime}} =  2_{1,1}\rightarrow 1_{1,0}}$~E line, was
detected towards 50 per cent of sources. Individual line profiles are
generally well fit by single Gaussians, and the average linewidth is
4.7\,\kms. The \chthreeoh~line profiles of five sources (G0.21+0.00,
G0.26+0.01, G6.54$-$0.11, G10.30$-$0.15, and G15.03$-$0.71) were too
confused or too weak to be successfully fit with Gaussians. The
parameters of the Gaussian fits to the remaining 74 spectra are
presented in Table~3.

\begin{figure}
  \begin{center}
    \includegraphics[width=8.0cm, angle=0]{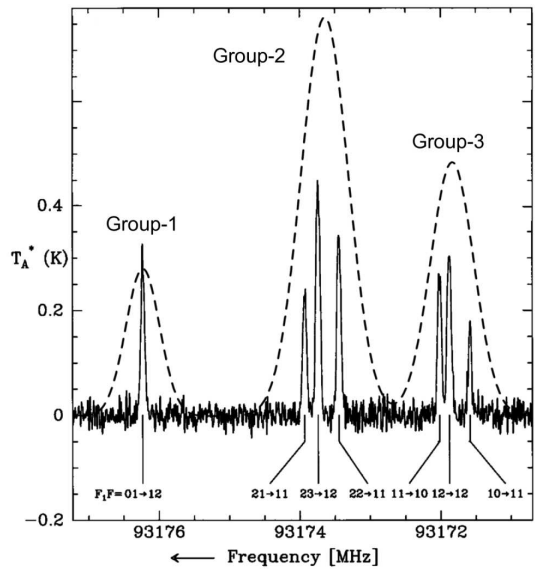}
    \caption[Spectrum of \ntwohp\,(1\,--\,0) towards a dark
      cloud.]{\small~Spectrum of \ntwohp\,(1\,--\,0) towards a dark
      cloud showing 
    the seven hyperfine components \citep{Caselli1995}. Linewidths
    towards massive star-forming regions are 4\,--\,5 times greater,
    resulting in the hyperfine components blending into three groups,
    designated 1\,--\,3 in this work (dashed lines).}
    \label{fig:n2hp_hyperfines}
  \end{center}
\end{figure}

The \ntwohp\,(1\,--\,0) transition is split into seven hyperfine lines, as
illustrated in Figure~\ref{fig:n2hp_hyperfines} (adapted from
\citealt{Caselli1995}). The relatively broad linewidths ($>$\,2\,\kms)
observed towards massive star forming regions result in the
seven components blending into three groups with roughly Gaussian
shapes. Consequently, we have characterised the profiles using two analysis
schemes. In the first instance we used the hyperfine structure fitting
(HFS) routine in the {\footnotesize CLASS} software to simultaneously fit the
\ntwohp~profiles with seven Gaussians. Free parameters were 
the velocity (V$_{\rm LSR}$), common linewidth, excitation
temperature, and optical depth. In the case of severely blended lines
(as presented in this work) the fits are not well constrained and
values for the optical depth and excitation temperature contain
substantial uncertainties. Using this method we find the mean
\ntwohp~linewidth of the sample is 3.0\,\kms. Because of the
uncertainty introduced by the blending, {\footnotesize CLASS} assigned an
erroneous value of $\tau\,=\,0.1$ to all sources, rendering the resultant
excitation temperatures meaningless. As an alternative measure of
optical depth and integrated intensity, we also fit the three blended
groups individually with single Gaussians. In total we
  successfully fit Gaussians to  79 spectra. The \ntwohp~emission
  in the source G0.26+0.01, G6.54$-$0.11 and G6.61$-$0.8 was too weak
  or two confused to achieve a reliable fit. We discuss the optical
depth and column densities derived from the \ntwohp~spectra in
Section~\ref{sec:n2hp_deriv}.  The results of the {\footnotesize CLASS} HFS
fits and the parameters of the Gaussian fits are presented in
Table~4.

Line profiles of HNC are observed to be similar to \hcop. The
critical densities of the two molecules are approximately the same at
$\sim 3\times 10^5$\,\cmmthree~\citep{Schilke1992} and it is likely
that they trace the same gas. Both molecules exhibit high velocity
line-wings. Self-absorbed HNC line profiles are common
in the sample and we referred to the low optical depth
\hthirteencop~and \ntwohp~line 
profiles when attempting to distinguish between self-absorption and
multiple clouds along the line of sight. We fit
self-absorbed profiles with a single Gaussian by masking off the
absorption dip (see Paper 1, Figure 3). We also measured
the line intensity by integrating under the line between velocity
limits bracketing the positive line emission and we calculated
an equivalent linewidth from 
\begin{equation}\label{eqn:ewidth}
  {\rm \Delta V_{equiv} = \frac{\int T_{mb}\,dv}{T_{mb}}.}
\end{equation}
In Equation~\ref{eqn:ewidth} ${\rm \int T_{mb}\,dv}$ is the integrated
intensity and T${\rm_{mb}}$ is the peak main-beam brightness
temperature. HNC emission from the sources G0.21+0.00 and
  G0.26+0.01 was not measured as the line profiles were too
  confused. In total 79 spectra were successfully fit with
  Gaussians, while 80 were measured by integrating under the line profile.
The results of the fits and integrated intensity measurements are
presented in Table~5.

The \thirteenco~line profiles often exhibit multiple components
spread over the 64\,MHz bandpass. Usually a single bright component
is present at the systemic velocity of the source, which is
easily fit with a single Gaussian. Occasionally several smaller
components are crowded around the main line and these are fit
as blended Gaussian lines. In general, the \thirteenco~profiles appear
moderately optically thick but exhibit less self-absorption than HNC
or \hcop. We again measure the integrated intensity and peak
brightness temperature directly from the raw spectrum. The parameters
of the fits and integrated intensity measurement of all 82
spectra are presented in Table~6.

\begin{figure*}
  \begin{center}
    \includegraphics[width=13.5cm, angle=0]{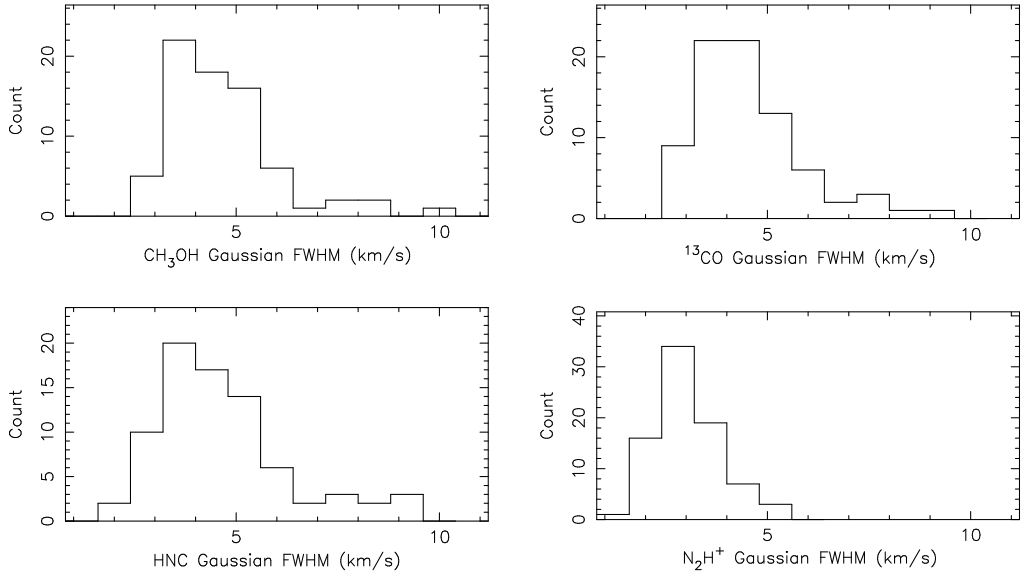}
    \caption[Distributions of full-width half-maximum linewidths.]{\small~Distributions of full-width half-maximum linewidths from
    the Gaussian fits to the spectra. The HNC and \thirteenco~line
    profiles were fit with single Gaussian components. \chthreeoh~was
    fit with multiple Gaussians, where the linewidths were constrained
    to have the same value. The \ntwohp~spectrum was fit using the
    {\small HFS} method in {\footnotesize CLASS}. Seven Gaussian components
    with a common linewidth were fit to the three groups of blended
    lines. The value plotted here is the linewidth of an
    individual component. The mean \chthreeoh, HNC and
    \thirteenco~linewidth are all at approximately 4.5\,\kms, while
    the mean \ntwohp~linewidth is at 3.0\,\kms.}
    \label{fig:linewidths}
  \end{center}
\end{figure*}
HCN exhibited highly confused line profiles, with few sources
exhibiting the expected 3-component hyperfine structure. We attribute
the complex profiles to a combination of high optical depths, multiple
clouds along the line of sight and low level emission from broad line-wings
tracing outflows. It is possible that some spectra may be contaminated by
emission from the reference position, as HCN appears to be ubiquitous near
the Galactic plane.

Like \hcop, high velocity outflows wings are common in the spectra of
HNC and, to a lesser extent, \thirteenco. We fit broad Gaussians to
these wings simultaneously with the main component, and their
parameters are presented in Tables~5
and~6 marked with a `w'. The Gaussian parameters for
blended lines are marked with a `b' in the same tables.


\subsection{Linewidths}
Figure~\ref{fig:linewidths} shows the distributions of full-width
half-maximum linewidths for the four molecules which were fit with
Gaussians. The median \chthreeoh, HNC and \thirteenco\, linewidths are all
at approximately 4.5\,\kms, while the median \ntwohp~linewidth is
uniformly lower at 3.0\,\kms. Kolmogorov-Smirnov (KS) tests yield
probabilities above 50 per cent 
that the \chthreeoh, HNC and \thirteenco~linewidth distributions are similar,
implying that these lines are emitted from gas with similar dynamics. In
contrast, the \ntwohp~distribution is signifficantly different, yielding
probabilities of less than 0.0001 per cent in each case. 

For a temperature of 58\,K (see Section \ref{sec:masses}) the median
intrinsic thermal linewidths for the 
molecules are as follows: \chthreeoh: 0.29\,\kms, \thirteenco: 0.31\,\kms,
HNC: 0.32\,\kms~and \ntwohp: 0.31\,\kms. These values represent only 7 to
10 per cent of the measured linewidths. Non-thermal line broadening
may arise from bulk motions, rotation or turbulence, which are
difficult to distinguise at the resolutions of these observations. We
can conclude that the \ntwohp~emission emanates from a more quiescent
region than the other molecules.


\section{Derived parameters}
We present the general physical parameters of the sources and
the excitation temperatures, column densities and optical depths
derived from the line profiles. The frequencies and electronic
constants used in the calculations are collected in
Table~\ref{tab:transitions}.


\subsection{Gas mass and virial mass}\label{sec:masses}
Gas masses have been calculated from the 1.2-mm continuum flux density,
 taken
from the work of \citet{Hill2005}. For a 1.2-mm integrated continuum
flux density S$_{\rm \nu}$ the mass of gas is given by \citet{Hildebrand1983}
\begin{equation}\label{eqn:dust_mass}
  {\rm M_{gas}=\frac{S_{\nu}\,D^2}{\kappa_d\,R_d\,B_v(T_{dust})},}
\end{equation} 
where D is the distance to the source in metres, ${\kappa_d}$ is the
mass absorption coefficient per unit mass of dust, B$_{\nu}$ is the
Planck function for a blackbody of temperature T$_{\rm dust}$, and
R${\rm _d}$ is the dust to gas mass ratio. We have adopted a value of 0.1
m$^2$\,kg$^{-1}$ for ${\rm   \kappa_d}$ at 1.2-mm and have assumed a
dust to gas mass ratio of 0.01 \citep{Ossenkopf1994}. The distance has been
estimated from the V$_{\rm LSR}$ of the source using the Galactic rotation
curve of \citet{BrandBlitz1993} (see Paper 1). Values for T$_{\rm 
  dust}$ were taken from the cold component of the greybody-fit to the
spectral energy distribution (SED) presented in Paper 1. For sources
where we had insufficient data to perform a fit we adopted the average
value of T$_{\rm   dust}$\,=\,58\,K, determined from 65 good fits. We
note that these temperatures are not well constrained by the sparsely
sampled SED and are generally higher than the values of 20\,--\,30\,K
adopted in the literature. For dust temperatures of 20\,K and 30\,K
the gas-mass derived here will be under-estimated on average by
factors of 3.6 and 2.1, respectively. Values for M$_{\rm gas}$ were
derived for 66 sources and are recorded in column six of
Table~7.

We estimated virial masses from the  \chthreeoh, \ntwohp,
\thirteenco~and HNC line-profiles following the approach by
\citet{MacLaren1988}. In the simplest case, neglecting support by
magnetic fields or internal heating sources, the total mass of a
simple spherical system is given by
\begin{equation}\label{eqn:mvirial_r2}
  {\rm M_{vir} = \frac{\sigma^2\,r}{G} = 126\,r\,\Delta v^2~~~~~~\ldots~~~~~~in~(\msun),}
\end{equation}
where $\sigma$ is the full 3-dimensional velocity dispersion, r is the
dust-radius of the cloud in pc, $\Delta$v is the FWHM linewidth of the
molecular line in \kms~and G is the gravitational constant in
N\,m$^2$\,kg$^{-2}$. The virial mass is the minimum mass required in
order for a cloud to be gravitationally bound, i.e. the cloud is bound if ${\rm
  M_{gas}/M_{vir}>1}$. If magnetic field support is important the
virial mass may be overestimated by up to a factor of two
\citep{MacLaren1988}. Observationally, the linewidths may be
artificially broadened due to blending of multiple components along
the line of sight, or in optically thick lines, due to radiative
transfer effects. Enhanced linewidths will cause us to overestimate
the virial mass by an unknown amount and the values quoted here should
be considered upper limits. Columns 7\,--\,10 of
Table~7 present the virial masses derived from the four
molecules. 

\begin{figure*}
  \begin{center}
    \includegraphics[width=14.0cm, angle=0]{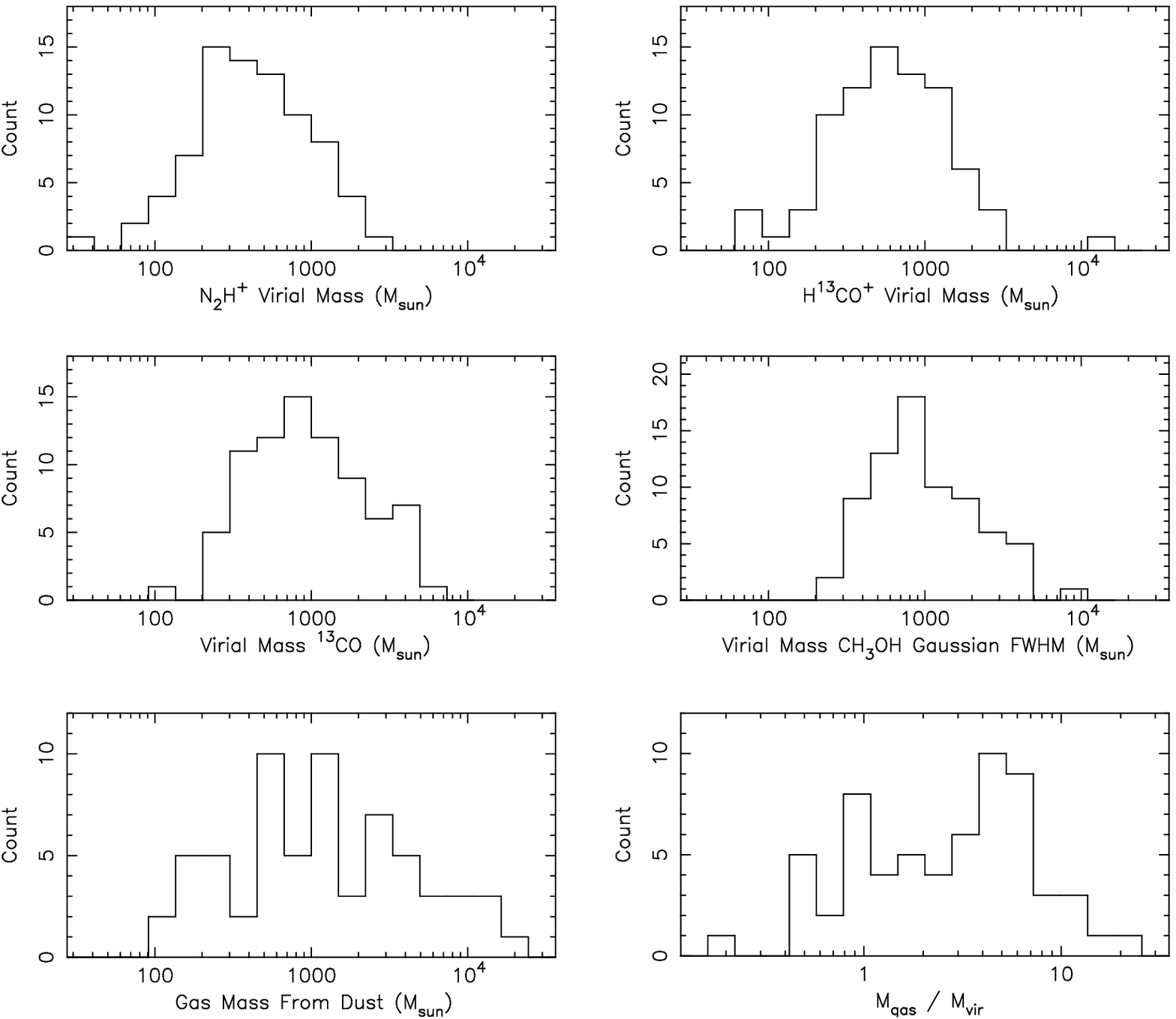}
    \caption[Distributions of virial
    masses.]{\small~Histograms showing the distributions of the virial
    masses derived from \ntwohp, \hthirteencop, \thirteenco~and
    \chthreeoh, and the gas masses derived from the 1.2-mm continuum
    emission from the work of \citet{Hill2005}. Virial masses should
    be considered upper limits. The plot on the bottom right shows the
    distribution of the ${\rm M_{gas}/M_{vir}}$ ratio derived from
    \ntwohp. The values of M$_{\rm gas}$ and M$_{\rm vir}$ are
    consistent with having the same order of magnitude, however,
    M$_{\rm vir,\,N_2H^+}$ $<$ M$_{\rm vir,\,others}$ as
    $\Delta$v$_{\rm N_2H^+}$ is $<$ $\Delta$v$_{\rm others}$.}
    \label{fig:hist_masses}
  \end{center}
\end{figure*}
The distributions of M$_{\rm gas}$ and M$_{\rm vir}$ are illustrated
in Figure~\ref{fig:hist_masses}. Optically thin \ntwohp~and
\hthirteencop~produce similar estimates of the mean virial mass at
573\,\msun~and 905\,\msun, respectively. \thirteenco~and
\chthreeoh~have the broadest linewidths, most likely due to
optical depth effects or low velocity outflows, leading to mean mass
estimates of 1293\,\msun~and  1305\,\msun, respectively. The values
derived for M$_{\rm gas}$ and M$_{\rm vir}$ are consistent with having
the same magnitude in the majority of clumps. The large uncertainties
associated with the calculations prevent the interpretation of
individual objects as gravitationally bound. 


\subsection{H$_2$ column and volume density}\label{sec:h2_densities}
We estimated the H$_2$ column
density, ${\rm N_{H_2}}$, from the gas-mass using the following relation:
\begin{equation}\label{eqn:N_h2}
  {\rm N_{H_2} = \frac{M_{gas}}{2\,M_{p}\,f_{He}\,A},}
\end{equation}
where ${\rm M_{p}\,=\,1.673\times 10^{-27}}$\,kg is the mass of a proton,
f$_{\rm He}\,=\,1.36$ is a correction factor to account for Helium present
in the interstellar medium \citep{Allen1973} and A is the projected
surface area of the mm-continuum emission. Assuming a cylindrical source with
uniform density, the projected area is given by
\begin{equation}
  {\rm A=\pi\,r^2\,D^2},
\end{equation}
where r is the angular radius of the source in radians
and D is the distance to the source in metres
(1\,pc\,=\,$3.08568\times 10^{16}$\,m). In a similar 
manner the volume density n$_{\rm H_2}$ can be estimated by replacing A in
Equation~\ref{eqn:N_h2} by V the projected volume of emitting gas
\begin{equation}\label{eqn:proj_vol}
  {\rm V=\frac{4}{3}\,\pi\,r^3\,D^3}.
\end{equation}
Equation~\ref{eqn:proj_vol} assumes emission from a uniformly
dense sphere of angular radius r. The radius of the dust continuum emission
was measured directly from the 1.2-mm continuum images (provided
courtesy of \citealt{Hill2005}) by fitting a Gaussian to the
azimuthally averaged source profiles. We find values for the
full-width half-maximum vary between 99\,\arcsec~and 24\,\arcsec~(the
limiting resolution of the SIMBA bolometer), and have a mean value of
34\,\arcsec. Derived values of N$_{\rm H_2}$ and n$_{\rm H_2}$ are
presented in columns four and five of Table~7 and have
means of 6.1$\times\,10^{22}$\,\cmmtwo~and
4.8$\times\,10^{4}$\,\cmmthree, respectively.

We urge caution in interpreting any value of the H$_2$ column density
derived in this manner as the assumptions introduce large unknown
errors. Firstly, the mass absorption coefficient and dust to
gas ratio used to calculate M$_{\rm gas}$ in
Equation~\ref{eqn:dust_mass} may vary considerably from source to
source, leading to a corresponding error in the H$_2$ column
density. The spatial distribution of H$_2$ is a further unknown
and additional errors are introduced by assuming a uniform column
density over a spherical projected volume. For the six sources at
  the limit of SIMBA resolution (24\,\arcsec) derived values should be
  considered beam averaged lower limits. Also see
  Section~\ref{sec:beam_dilution} for a discussion on beam dilution.


\subsection{\chthreeoh~rotational temperature and column density}
\chthreeoh~is a slightly asymmetric rotor, resembling a symmetric top
except for the O-H group, which is angled with respect to the
principal axis. Internal hindered motion
coupled with the rotation of the molecule results in a complicated
rotational spectrum, organised into non-degenerate A and two-fold
degenerate E levels, corresponding to different torsional
symmetry. The A and E states have nuclear statistical 
weights of 2 and 1, respectively, thus their overall degeneracy is the
same \citep{Bockelee1994}.

A rotational temperature ${\rm T_{rot}}$ and column density N may
be estimated from the ratios of the thermal \chthreeoh~lines based on the
assumptions of low optical depths and that the background temperature
(${\rm T_b\approx 2.7\,K}$) is small compared to the brightness
temperature of the line. Under conditions of Local Thermodynamic
Equilibrium {\it all} the observed energy levels can be described by a
single excitation temperature T$_{\rm ex}$, defined by the Boltzmann
relation. Using the relation for rotational temperature
\citep{GoldsmithLanger1999},
\begin{equation}
  {\rm ln\,\left(\frac{N_u}{g_u}\right) = ln\left(\frac{N}{Q(T)}\right) -
  \frac{E_u}{k\,T_{rot}}}
\end{equation}
a rotation diagram may be drawn by plotting ln\,(${\rm N_u/g_u}$)
versus ${\rm  E_u/k}$. The rotation temperature and column density in
the upper level ${\rm N_u}$ are found from the slope and y-axis
intercept, respectively, of a straight line fit to the data. The total
column density N may then be found assuming local thermal equilibrium
and a partition function Q(T$_{\rm rot}$), which for \chthreeoh~is
given by \citep{Townes1955} 
\begin{equation}\label{eqn:q_ch3oh}
  {\rm Q(T) = \sum_{J=0}^{\infty}(2J+1)\,e^{-E_J/kT}~\approx~1.2327\,T_{rot}^{1.5}.}
\end{equation}
In Equation~\ref{eqn:q_ch3oh} we have approximated the partition
function by a fit to the discrete values in the \citet{Pickett1998}
spectral line catalogue. Note that values for ${\rm E_u}$
are referenced to different ground state energy levels in A and
E-type \chthreeoh. The ground state for A-type is the ${\rm
  J_K}\,=\,0\,_0$ level, while for E-type \chthreeoh~is the  ${\rm
  J_K}\,=\,1\,_{-1}$ level \citep{Menten1988}. This is commonly
overlooked in the literature as all values quoted in
\citet{Pickett1998} are referenced to the ${\rm J_K}\,=\,0\,_0$ level,
resulting in incorrect rotation diagrams. The correct values of ${\rm
  E_u/k}$ are cited in Table~\ref{tab:transitions}. 

\begin{figure*}
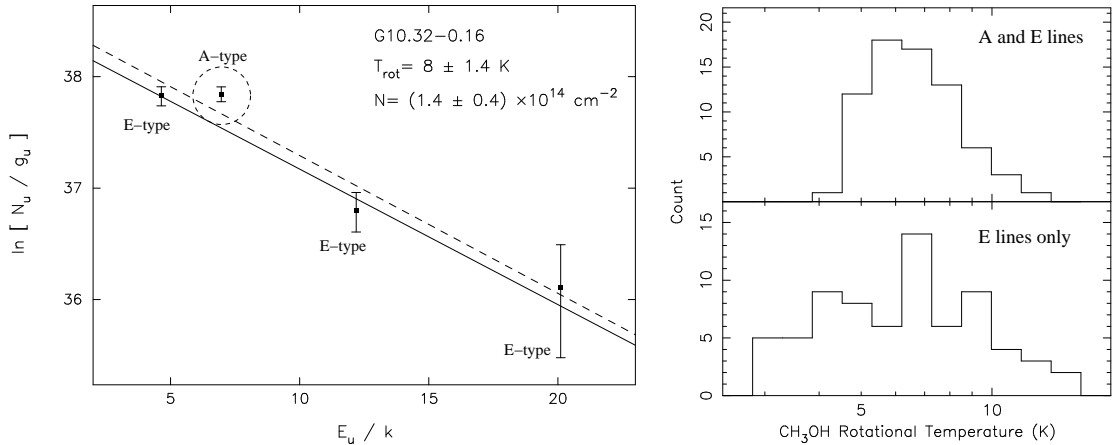

  \centering
  \includegraphics[height=6.2cm, angle=0, trim=0 -20 0 0]{figs/fig5a_ch3oh_rotdiag_eg.epsi}
  \includegraphics[height=6.2cm, angle=-00, trim=-20 -20 0 0, origin=c]{figs/fig5b_hist_ch3oh_trot.epsi}
  \caption[Rotational diagram of \chthreeoh~for
    the source G10.32$-$0.16.]{\small~({\it left}) Rotational diagram of \chthreeoh~for
    the source G10.32$-$0.16, made using the 96.7\,GHz lines. The
    point at ${\rm E_u} \approx\,7$\,K corresponds to the
    $2_0\rightarrow\,1_0$\,A+ transition. The three other points are
    due to transitions of E-type methanol. The solid line and values
    for ${\rm T_{rot}}$ and N correspond to a fit through the E
    transitions only. The dashed line represents a fit to all
    transitions. ({\it right}) Distributions of rotational
    temperatures derived from \chthreeoh~using both A and E-type
    transitions (top), and E-type transitions only (bottom).}
    \label{fig:ch3oh_rotdiag_eg}
\end{figure*}

An example of a \chthreeoh~rotation diagram is shown in
Figure~\ref{fig:ch3oh_rotdiag_eg} ({\it left}). The data point at
${\rm E_u}$/k\,$\approx$\,7\,K corresponds 
to the A+ line, while the other three points correspond to E
transitions. In general, where we detect three E-type lines, they are
well fit with a straight line, implying that our assumption of
optically thin conditions is valid.

A and E type methanol are oblate and prolate forms, respectively, of
the assymetric top and only inter-convert on timescales of $\sim$10,000
years. It is not clear that
the A/E ratio can be assumed equal to one and a rotation diagram
including both A and E type transitions may not be valid. In
Figure~\ref{fig:ch3oh_rotdiag_eg} ({\it left}) the solid line
indicates a fit to the E-type transitions only, while the dotted line
is a fit to both the A and E-type. Rotation temperatures derived from
only the E-transitions have a mean value of 6.67\,K, similar to the
mean temperature derived from all four transitions, which is
6.8\,K. Rotational temperatures and column densities derived using
both fitting methods are presented in Table~8 and
the distributions of rotational temperatures are illustrated in
Figure~\ref{fig:ch3oh_rotdiag_eg} ({\it right}).

Column densities are averaged over the telescope beam and are likely
underestimated. If the bulk of the \chthreeoh~emission stems from
within a hot core then the typical size of the emitting region will
be $\sim$0.1\,pc ($\approx 20,000$\,AU, \citealt{Kurtz2000}). Columns
three and four of Table~8 present the angular
size and beam dilution factor calculated using this
assumption. Columns seven and ten present the column densities
corrected for this assumed dilution factor.

On inspection of the bulk of the rotation diagrams (available as an
online supplement) we see a pattern emerge. The A-type
transition has a 
consistently higher ${\rm N_u/g_u}$ compared to a straight line fit
through the  E-type transitions. This indicates a greater abundance of
A-type \chthreeoh~and an A/E ratio $>$\,1. Considering only sources in
which all three E-type lines were detected (36 sources), and assuming equal
excitation temperatures in A and E lines, we derive A/E abundance
ratios ranging from 0.47 to 1.80, with a mean of 1.30. These values are
comparable to values previously derived in the literature, e.g.,
\citet{Menten1988} who calculated A/E ratios between 1.3 and 2.0, using
the same transitions, towards warm clumps in Orion KL.


\subsection{\ntwohp~optical depth and column density}\label{sec:n2hp_deriv}
\begin{figure}
  \begin{center}
    \includegraphics[angle=-90, width=7.0cm, trim=0 0 0 0]{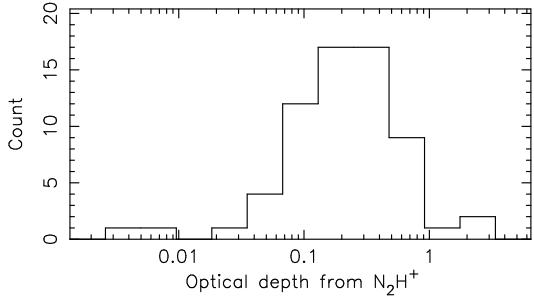}
    \caption{\small~Distribution of optical-depths derived from the 
    ratios of the hyperfine groups of \ntwohp. The mean is
    0.35\,$\pm$\,0.42.}
    \label{fig:hist_n2hp_tau}
  \end{center}
\end{figure}

Optical depths were derived from the blended hyperfine components
of \ntwohp, which assume set ratios under optically thin conditions
(see Section~\ref{sec:line_profile_params}). Assuming the linewidths
of the individual hyperfine components are all equal, the integrated
intensities of the three blended groups should be in the ratio of
1\,:\,5\,:\,2 under optically thin conditions. The optical depth is
calculated from the ratio of integrated intensities (${\rm
  \int\,T_{B}\,dv}$) of any two groups using the following equation: 
\begin{equation}
  {\rm \frac{\int\,T_{B,1}\,dv}{\int\,T_{B,2}\,dv} = 
  \frac{1-e^{-\tau_{1}}}{1-e^{-\tau_{2}}} =
  \frac{1-e^{-\tau_{1}}}{1-e^{-a\,\tau_{1}}},}
\end{equation}
where `$a$' is the expected ratio of ${\rm \tau_{\,2}/\tau_{\,1}}$
under optically thin conditions. \citet{Caselli1995} report anomalous
excitation of the ${\rm F_1,F=1,0\rightarrow 1,1}$ and
${1,2\rightarrow 1,2}$ components (in our group-3), so we determine
the optical depth solely from the intensity ratio of
group-1/group-2. We find that 97 percent of sources have optical
depths below 1. Figure~\ref{fig:hist_n2hp_tau} plots the distribution
of the derived optical depths, which range from 0.035 to 2.43, with a
mean of 0.35 and a standard-deviation of 0.42. Values for the optical
depth towards individual sources are presented in column two of
Table~9. 

Column densities of \ntwohp~can be found from the following standard
formula \citep{Rohlfs2004},
\begin{equation}\label{eqn:n2hp_col}
{\rm N = \frac{8k\pi\nu^2\,\int T_b\,dv}{A_{ul}hc^3}\left(\frac{\tau_{\nu}}{1-e^{-\tau_{\nu}}}\right)\frac{Q(T_{ex})}{g_u}\,e^{E_u/kT},}
\end{equation}
where Q(T$_{\rm ex}$) is the partition function, extrapolated from a
fit to the values given by the \citep{Pickett1998} spectral line
catalogue: 
\begin{equation}
  {\rm Q(T)= 4.198\,T_{\rm ex}.}
\end{equation}
Equation~\ref{eqn:n2hp_col} requires an estimate of the excitation
temperature T$_{\rm ex}$. Unfortunately, under optically
thin conditions, the excitation temperature cannot be determined from
the \ntwohp~line profiles, so we assume a constant value of 10\,K. For
excitation temperatures of 5\,K and 20\,K the total column density
should be multiplied by 0.8 and 1.6, respectively. Beam averaged
column densities derived using Equation~\ref{eqn:n2hp_col} are
presented in column five of Table~9. In the
literature, high-resolution maps show that \ntwohp~traces the spatial
extent of the dust closely
(e.g. \citealt{Caselli2002a,Pirogov2003}). We have estimated the
average FWHM of the dust emission for all sources observed by
\citet{Hill2005} (see section~\ref{sec:h2_densities}) and use these
values to correct for beam dilution. Total column
densities corrected by a factor ${\rm
  (\theta_{dust}/\theta_{beam})^2}$ are presented in column six of
Table~9.


\subsection{\thirteenco, HNC and HCN column density}
The HNC, HCN and \thirteenco~profiles frequently display asymmetries which
may indicate that they are optically thick in general. Lacking an
estimate for both the excitation temperature and the optical depth we
are unable to calculate reliable column densities.


\subsection{Abundances}\label{sec:hmc_2_abundance}
The relative abundance X between two species may be found directly
from the ratio of their volume densities. Assuming both molecules
occupy the same volume of space, this is simply equivalent to the
ratio of their column densities.
\begin{equation}\label{eqn:xfactor}
  {\rm X=\frac{n_{1}}{n_{2}} \approx \frac{N_{1}}{N_{2}}.}
\end{equation}
If, however, one or both species subtend angles smaller than the beam,
we must correct for their {\it relative} volume filling factor and
Equation~\ref{eqn:xfactor} becomes
\begin{equation}
  {\rm X=\frac{N_{1}}{N_{2}}\,\left(\frac{\theta_2}{\theta_1}\right)^3,}
\end{equation}
assuming spherically symmetric emission with characteristic sizes
$\theta_1$ and $\theta_2$. For the single-position observations presented
here, the volume filling factors are complete unknowns and constitute
the greatest source of errors in the abundance. Additional errors stem from
calibration of the brightness scale ($\sim$\,30 per cent) and the
pointing accuracy of the telescope ($\sim$\,20 per cent). Abundances
relative to H$_2$ depend also on the assumptions made in calculating
the gas mass from the 1.2-mm continuum emission and cannot be
determined reliably.


\section{Luminosities and masses}\label{sec:hmc_lum_mass}
\begin{figure*}
  \begin{center}
    \includegraphics[origin=c, width=5cm, angle=-90, trim=0 0 100 0]{figs/fig7a_xy_lum_mass.epsi}   
    \includegraphics[height=5cm, angle=0, trim=-10 -20 0 70]{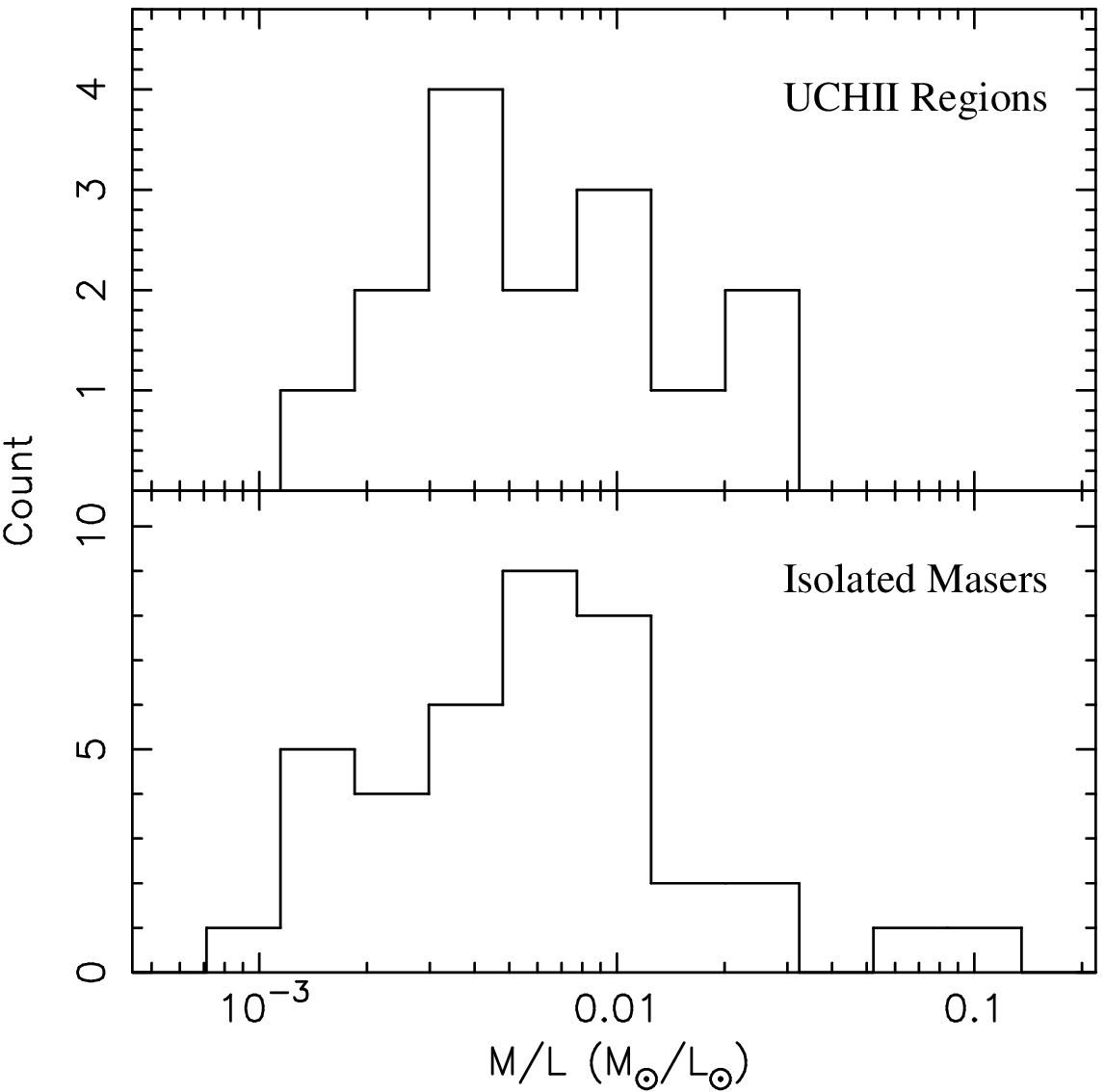} 
    \caption[Plot of luminosity versus dust\,+\,gas mass.]{\small~({\it left}) Plot of luminosity versus dust\,+\,gas mass for 55
    sources whose properties could be determined (after Fig.~2,
    \citealt{Cesaroni2005}). \uchii~regions are represented by solid
    triangles and isolated masers by crosses. The thick
    solid line traces the luminosity expected if ${\rm M_{gas}\equiv
    M_{stars}}$ in the clumps, and the stellar population is
    concentrated in a minimum number of stars with masses 
    $\leq$100\,M$_{\sun}$. The dashed line is the luminosity expected
    from the most massive star in a M$_{\rm gas}$ cluster with
    stellar masses distributed according a Salpeter IMF
    ($\alpha=2.35$), and a fundamental upper stellar mass limit of
    150\,M$_{\sun}$ \citep{Weidner2004}. We have assumed a
    mass-luminosity relation of ${\rm L\,\propto\,M^{3}}$. The
    dash-dot line traces the expected luminosity from the same IMF
    assuming ${\rm L\,\propto\,M^{4}}$ and the shaded region indicates
    intermediate values. The observed mass-luminosity relation is fit
    with a single power law of index 0.68\,$\pm$\,0.09 (thin solid
    line). ({\it right}) Histograms of the distribution in the
    M$_{\rm gas}$/L$_{\rm stars}$~values for \uchii~regions and
    isolated masers. A KS-test cannot distinguish between them.}
    \label{fig:xy_mass_lum}
  \end{center}
\end{figure*}

The relationship between mass and luminosity provides clues to the
star formation efficiency, stellar content and evolutionary state of
the clumps. Figure~\ref{fig:xy_mass_lum} ({\it left}) is an expanded
version of Fig.~2 in \citet{Cesaroni2005}, showing the relationship
between luminosity and gas mass for the 55 clumps whose properties
could be determined. The luminosity has been measured from a
two-component greybody fit to the spectral energy distribution and is directly
related to the stellar content of the clump. Sub-millimetre emission
from assumed optically thin dust \citep{Hill2005} has been used to
derive the total gas-mass of the clumps. In the plot, the thick line
marks the expected luminosity from a minimum number of stars of mass
M$_{\ast}\leq$100\,M$_{\sun}$, assuming ${\rm M_{gas}\equiv
  M_{stars}}$. \citet{Cesaroni2005} divide a sample of 21 known hot
cores\footnote{We have the following sources in common with Cesaroni:
  G5.89$-$0.39, G9.62+0.19, G10.47+0.03, G10.62$-$0.38, G29.96$-$0.02
  and G31.41+0.31.} into `light' and `heavy' categories, based on
their position above or below this division. `Light' cores have gas
masses below $\sim$100\,M$_{\sun}$ {\it and} fall above the thick
line. Their high luminosity-to-mass ratio implies that the mass of their stellar
content is greater than that of the gas. `Heavy' cores have gas masses
above $\sim$100\,M$_{\sun}$ and fall below the thick line. The mass of
their stellar content is comparable to, or less than, the mass of the
gas. \citet{Cesaroni2005} have interpreted the L$_{\rm stars}$/M$_{\rm
  gas}$ ratio in the `light' cores as suggesting only a single
embedded massive star exists in these clumps. Conversely the `heavy'
clumps are likely to contain multiple massive stars which have formed
in clustered mode. We do not find any `light' cores in our sample.

An alternative explanation for high L$_{\rm stars}$/M$_{\rm gas}$
ratios in the `light' sources might derive from the age of the star
forming regions. A massive young stellar object at a late stage of
evolution could be expected to have dispersed a large fraction of its
molecular gas, through the action of stellar winds, intense
UV-radiation and outflows from young OB stars. This would
result in a 
decrease in the measured gas-mass without a corresponding decrease in
the luminosity. In Figure~\ref{fig:xy_mass_lum} the points would move
horizontally into the `light' portion of the plot. We note that
Cesaroni's light cores include Orion-KL, and IRAS\,20126+4014, neither
of which are believed to be at a late stage of evolution. Orion is the
closest massive star forming region and is not clear if it is
representative of massive star-forming regions in general. The
mass of IRAS\,20126+4014 (7\,$\pm$\,3\msun) has been measured
assuming Keplerian rotation of a molecular disk
\citep{Cesaroni2005a} and relates to the inner object only. On larger
scales the gas-mass has been measured to be several hundred solar
masses (e.g. \citealt{Estalella1993}), putting IRAS\,20126+4014 in the
`heavy' category. We find that all of our maser selected sample have
gas masses above $\sim$30\,M$_{\sun}$ and {\it all} fall into the
`heavy' category. If the division between `light' and `heavy' cores is
an indication of age, then all of our sample are at a relatively early
stage of evolution, i.e., the powering massive young stellar objects
have not yet dispersed their natal molecular clouds.

In Figure~\ref{fig:xy_mass_lum} the shaded region between the dashed
lines represents the range of luminosities expected from the {\it most
  massive} star in a cluster where ${\rm M_{stars}=M_{gas}}$, if the
initial mass function (IMF) follows a Salpeter power law ($\alpha
=2.35$) and assuming a fundamental upper stellar mass limit of ${\rm
  M_{\ast} =   150\,M_{\sun}}$ \citep{Weidner2004}. The luminosity of
the cluster will be dominated by the most massive star so this is a
good approximation to the expected total luminosity. The range in
expected  luminosities derives from the uncertainty in the
mass-luminosity relation for massive young stellar objects. This may
be represented by a power law  of the form ${\rm L_{\ast}\propto
  M_{\ast}^{\gamma}}$, where $\gamma$ ranges between 3 and 4 in the
literature for OB-stars (e.g., \citealt{Lada1999,Churchwell2002}). Our
maser selected sample exhibit a rough correlation between mass and
luminosity, which may be fit with a single power law with index
$\gamma = 0.68\pm 0.08$. This is a much shallower rise than observed
in evolved stars, however, the luminosities of the most massive
clusters are consistent with the 150\,M$_{\sun}$ stellar cutoff
\citep{Weidner2004}, even in the more conservative limit of ${\rm
  L_{\ast}\propto M_{\ast}^{3}}$.  
\begin{figure*}
  \begin{center}
    \includegraphics[width=13cm, angle=0, trim=0 0 -20 0]{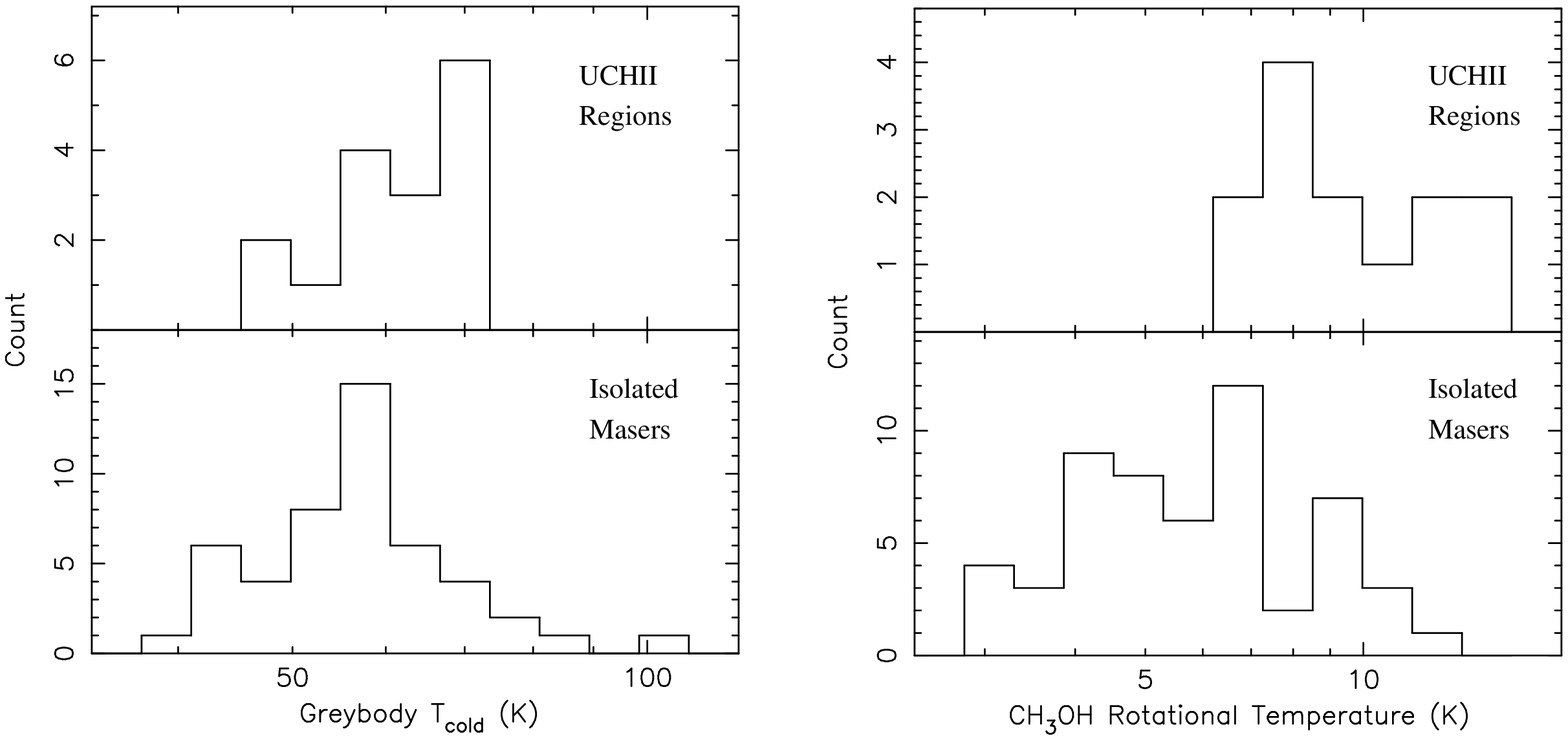}
    \caption[Distribution of cold-component dust
    temperatures.]{\small~({\it left}) Histograms showing the
    distributions of 
    dust temperature taken from the cold component of the greybody fit
    to the SED. The top panel shows \uchii~regions, which have a mean
    T$_{\rm cold}$ of 62.4\,K, while the bottom panel shows isolated
    masers, which have a mean T$_{\rm cold}$ of 58.1\,K. ({\it right})
    Histograms showing the distributions of \chthreeoh~rotational
    temperature for \uchii~regions (top) and isolated masers
    (bottom). The mean values are 9.9\,K for \uchii~regions and 6.1\,K
    for isolated masers. A KS-test establishes that the
    \chthreeoh~rotational temperature distributions are significantly
    different, returning a probability of 0.002 per cent.}
    \label{fig:hist_tdust}
  \end{center}
\end{figure*}

The derived gas-mass is highly
dependent on the temperature in Equation~\ref{eqn:dust_mass}. The
shallow power law might be explained if the average dust temperature
increases with luminosity and this change is not reflected in the
sparsely sampled greybody fits to the SED, from which we take our
temperature estimates. If we assume the mass estimates are correct to
within a factor of a few, then the main result is that the lower mass
clumps are over-luminous and the higher mass clumps are
under-luminous. 


\subsection{\uchii~regions: Luminosity or Age?}
\uchii~regions (triangles) are in general more luminous {\it and} more
massive than the isolated masers (crosses) in the sample. Individual
power law fits to the isolated masers and \uchii~sub-samples yield
mass-luminosity relations of ${\rm L_{bol\,|\,masers}\propto M^{0.59\pm
    0.13}}$ and ${\rm L_{bol\,|\,UCHII}\propto M^{0.64\pm 0.18}}$,
respectively. These fits are consistent with the fit to the combined
sample, within the errors, and a single power law can
describe both distributions. Figure~\ref{fig:xy_mass_lum} ({\it
 right}) is a histogram of the M$_{\rm gas}$/L$_{\rm stars}$ ratio for
\uchii~regions (top) and isolated masers (bottom). As expected from
the mass-luminosity plot, the distributions are similar and a
 KS-test cannot distinguish between them, yielding
a 76 per cent probability of the two sub-samples being drawn from the same
population. This result implies that for the same masses, the two
samples have the same luminosities. Early in the evolution of a
massive young stellar object we would expect its luminosity to
increase as it accretes matter and more mass is transfered from the
envelope to the star. At the same time the overall mass of gas and
dust in the clump is expected to remain relatively constant, or indeed
decrease, as winds from the young OB-cluster break up the cloud. By this 
rationale, if our sample of \uchii~regions are older and more evolved,
we would expect them to have a lower M$_{\rm gas}$/L$_{\rm stars}$
ratio than the isolated masers. There are several possible reasons why
we do not see a difference between the mass-luminosity distributions.

It is possible that systematic errors (which we discuss below)
associated with our assumptions 
cause us to overestimate the mass of the clouds hosting the
\uchii~regions relative to the clouds hosting the isolated masers. For
there to be no difference between the mean masses of the sub-samples,
the isolated masers would need to have a dust temperature on average
36\,K lower than the 
\uchii~regions. Figure~\ref{fig:hist_tdust} ({\it left}) illustrates
the distributions of T$_{\rm cold}$ from the SED fits, for
\uchii~regions (top) and isolated masers (bottom). While the
\uchii~regions clearly have temperatures which are skewed to the high
end, the difference in the means is only 4.3\,K. The rotational 
temperatures calculated from \chthreecn~are likewise similar, however,
we cannot rule out different average dust temperatures on the basis
that \chthreecn~emission likely probes only the embedded `hot core'
and does not reflect the conditions in the extended
clump. Thermal \chthreeoh~emission is thought to derive from a more extended
envelope and rotational temperatures calculated from this molecule appear
to support very different temperatures in the two
groups. Figure~\ref{fig:hist_tdust} ({\it right}) shows histograms of 
the \chthreeoh~rotational temperature for the \uchii~regions (top) and
isolated masers (bottom). A KS-test returns a probability of 0.002 per
cent of the distributions deriving from the same parent. We note that
the mean values of 9.9\,K for the \uchii~regions and 6.1\,K for the
isolated masers do not reflect the kinetic temperature as the
\chthreeoh~is likely sub-thermally excited (see
\citealt{GoldsmithLanger1999}).
\begin{figure*}
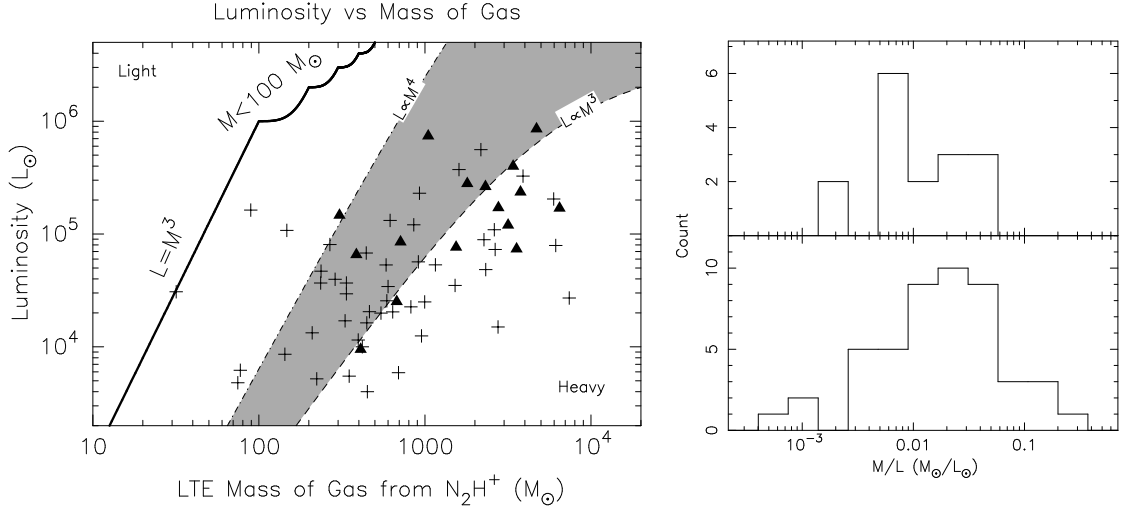

  \begin{center}
    \includegraphics[width=6.8cm, angle=-90,  trim=0 0 -05 0]{figs/fig9a_xy_lum_ltemass.ps}
    \includegraphics[width=6.8cm, angle=-90,  trim=-28 -20 -25 0]{figs/fig9b_hist_ltemass_lum.ps}
    \caption{\small~As Figure~\ref{fig:xy_mass_lum}, except in this
    case the mass is the LTE-mass derived from the column density of
    \ntwohp~assuming a fixed abundance ratio of
    $1.7\,\times\,10^{-10}$ and that the \ntwohp~emission closely
    follows the spacial distribution of the 1.2-mm continuum emission.}
    \label{fig:xy_ltemass_lum}
  \end{center}
\end{figure*}

As an alternate estimate of the mass we derived a `LTE-mass' from
the column density of \ntwohp. In our calculations we assume (i) the
\ntwohp~emission closely traces the 1.2-mm dust continuum and has a
flat density distribution within the FWHM of the clump, (ii) a fixed [H$_2$]\,/\,[\ntwohp] abundance ratio of
$1.7\,\times\,10^{-10}$ \citep{Pirogov2003}. The mass of gas may then
be estimated from the column density using
Equation~\ref{eqn:N_h2}. Figure~\ref{fig:xy_ltemass_lum} ({\it left})
is is a mass-luminosity plot similar to Figure~\ref{fig:xy_mass_lum},
except that the mass plotted on the x-axis is the LTE-mass calculated
from \ntwohp. Although the scatter is greater, we clearly see
the masses of the \uchii~regions and isolated masers are still weakly
offset, with the 
\uchii~regions having greater masses on average, as in
Figure~\ref{fig:xy_mass_lum}. The distributions of the M$_{\rm
  gas}$/L$_{\rm stars}$ ratios are illustrated in
Figure~\ref{fig:xy_ltemass_lum} ({\it right}). A KS-test returns
a probability of 50 per cent that both distributions are derived
from the same parent, confirming the original result.

Another explanation for the offset in the masses derives from the
sensitivity limit on the radio-survey. In the preceeding paper we
assumed the \uchii~regions were older and more evolved than the
isolated masers. We argued that the original radio continuum survey
was sensitive enough to detect \uchii~regions associated with the less
luminous isolated masers (see Paper 1). However, if a
sensativity bias existed it would manifest as a mass and luminosity
offset between the radio-loud and radio-quiet, i.e, the \uchii~regions
would be {\it more massive and more luminous} than 
isolated masers. In other words, some of the lower L$_{\rm
  stars}$ isolated masers may indeed be associated with \uchii~regions which
fall below our detection  limit. In Paper 1 we showed
that most sources are luminous enough to have detectable
\uchii~regions, assuming up to 90 per cent of the Lyman continuum
photons are absorbed by dust \citep{KurtzChurchwellWood1994}. If the
disparity in luminosity is due to a real sensitivity limit, this would
imply that, on average, at least 99 per cent of the ionising photons
are absorbed by dust before contributing to the creation of a
\uchii~region. We believe this is unlikely in the majority of
  cases.

Some of the isolated maser sites may also be associated with
hyper-compact \hii~(HCH{\scriptsize II}) regions. This recently
discovered class of \hii~region is typically less than 0.01\,pc
across \citep{Alvos2006} and exhibits a free-free emission spectrum
with a turnover at higher frequencies than \uchii~regions,
typically 20\,GHz. HCH{\scriptsize II} regions are expected to be
optically thick at 8.6\,GHz due to the squared dependace of the
flux density with frequency and will be difficult to
detect. \citet{Longmore2007} had used the Australia Telescope
Compact Array to search for 22\,GHz emission towards 24 of the
objects presented here. Nine of these sources fall into the radio-quiet
category and of these seven exhibit compact (30\,$\rightarrow$\,3\arcsec)
continuum emission. This result strongly suggests that some of our
radio quiet sources are associated with HCH{\scriptsize II} regions,
or indeed \uchii~regions undetected at 8.6\,GHz. However, because
HCH{\scriptsize II} regions are thought to be the precursors of
\uchii~regions, we are still be selecting for two different
evolutionary states.

The most obvious explanation for the offset is due to the
timescales on which we expect massive young stellar objects to
progress to develop a detectable \uchii~region. In essence, clumps
containing very high mass objects will evolve very
rapidly and will spend only a short time in hot core and earlier
phases. Because of the short timescales involved, very massive hot
cores will also likely be deeply embedded. The first indication of
star-formation activity may be when the expanding \hii~region begins
to disperse the surrounding molecular gas. Anecdotal evidence
supporting this explanation is already emerging from uniform
searches for MYSOs, such as the Red MSX Source (RMS) survey
(\citealt{Hoare2004}, \citealt{Mottram2008}). The
less evolved MYSOs identified in the survey have a significantly
smaller upper-mass-cutoff compared to the \uchii~regions
detected. No MYSO counterparts are found to the most massive
\uchii~regions, implying that the \uchii~phase is the first
indication of star-formation in the most massive clumps.

{\it We believe differences between the radio-bright and radio-quiet
  samples may be attributed to evolution, with the caveat that the
  highest luminosity objects do not spend significant time in an
  `isolated maser' phase.}


\section{Detailed comparison between sub-samples}\label{sec:detail_comp}
\begin{figure*}
  \begin{minipage}[t]{8.5cm}
    \centering

    \includegraphics[height=7cm, angle=-90, trim=0 0 -10 0]{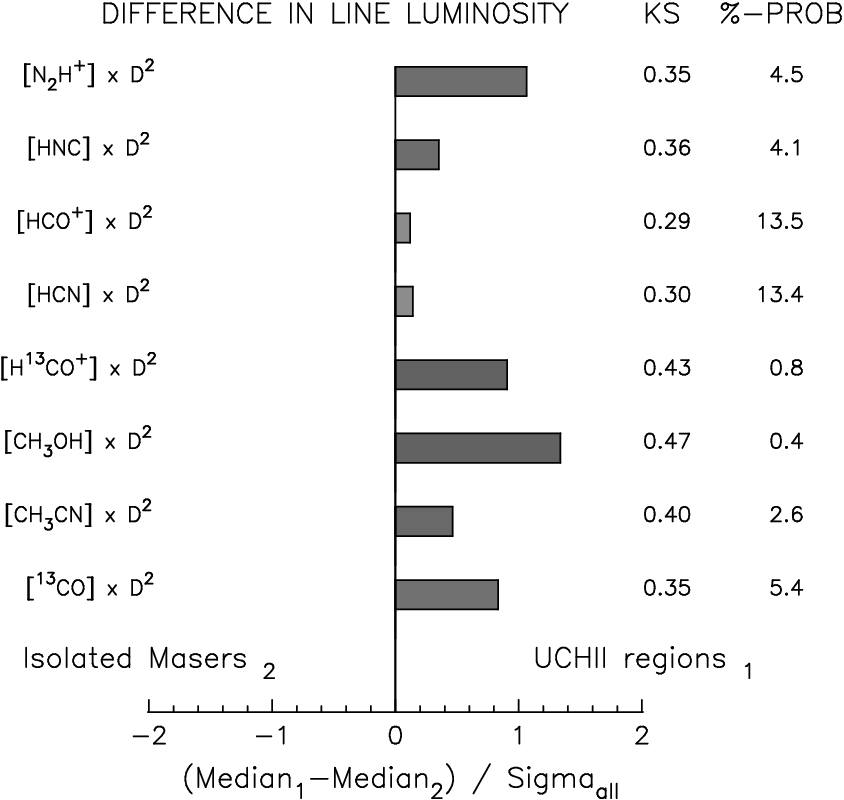}   	
    \caption{\small~The bars in this horizontal histogram represent the
    difference between the median values of two distributions,
    normalised to the standard deviation of the combined sample. Here, we
    compare {\it spectral line luminosity} in sources with and without
    \uchii~regions. A positive value means the line is more luminous
    towards \uchii~regions. To assess the significance of the
    differences, we compute the KS-statistic (left column) and
    associated percentage probability (right column). As a visual aid,
    the greyscale of the bars is scaled to the probability. Probabilities
    smaller than approximately two per cent (dark bars) indicate
    significant differences.}
    \label{fig:diffm_intd2_uchii}
    \end{minipage}
  \hfill
    \begin{minipage}[t]{8.5cm}
      \centering
      \includegraphics[height=7cm, angle=-90, trim=0 0 -5 0]{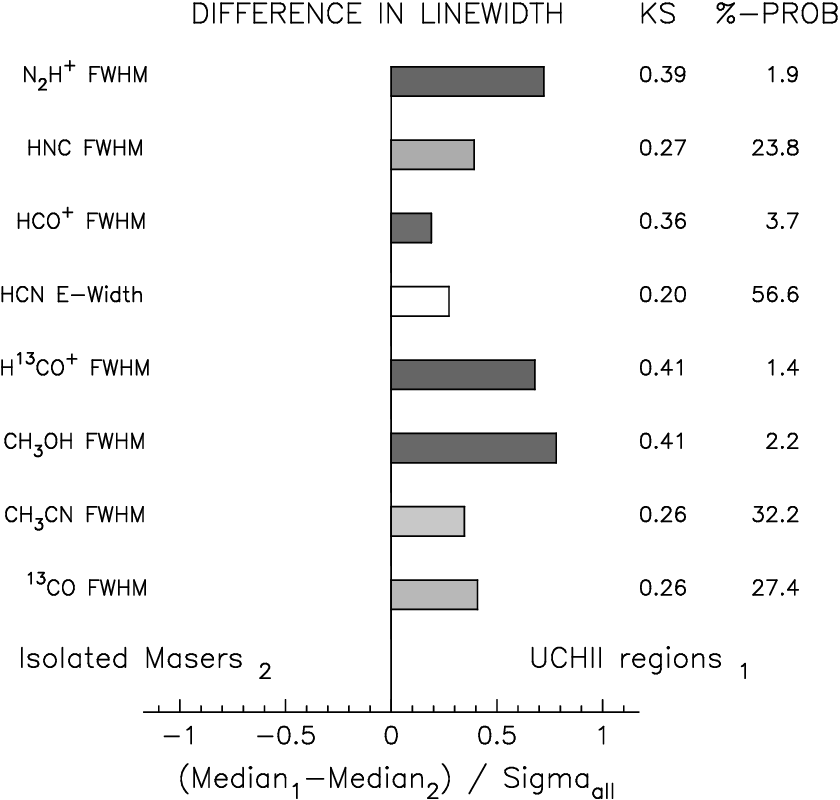}
	\caption{\small~Differences in the median linewidth, for
	  sources with and without an associated \uchii~region. The
	  linewidths were measured in two ways: by fitting a Gaussian to
	  the line profile and measuring the full-width at half-maximum, or
	  by integrating under the line and computing the equivalent
	  width (HCN only). Other details are as in Figure \ref{fig:diffm_intd2_uchii}.}
	\centering
	\label{fig:diffm_lw_uchii}
    \end{minipage}
\end{figure*}
\begin{figure*}
  \begin{minipage}[t]{8.5cm}
    \begin{center}
      \includegraphics[height=6.8cm, angle=-90, trim=0 0 0 0]{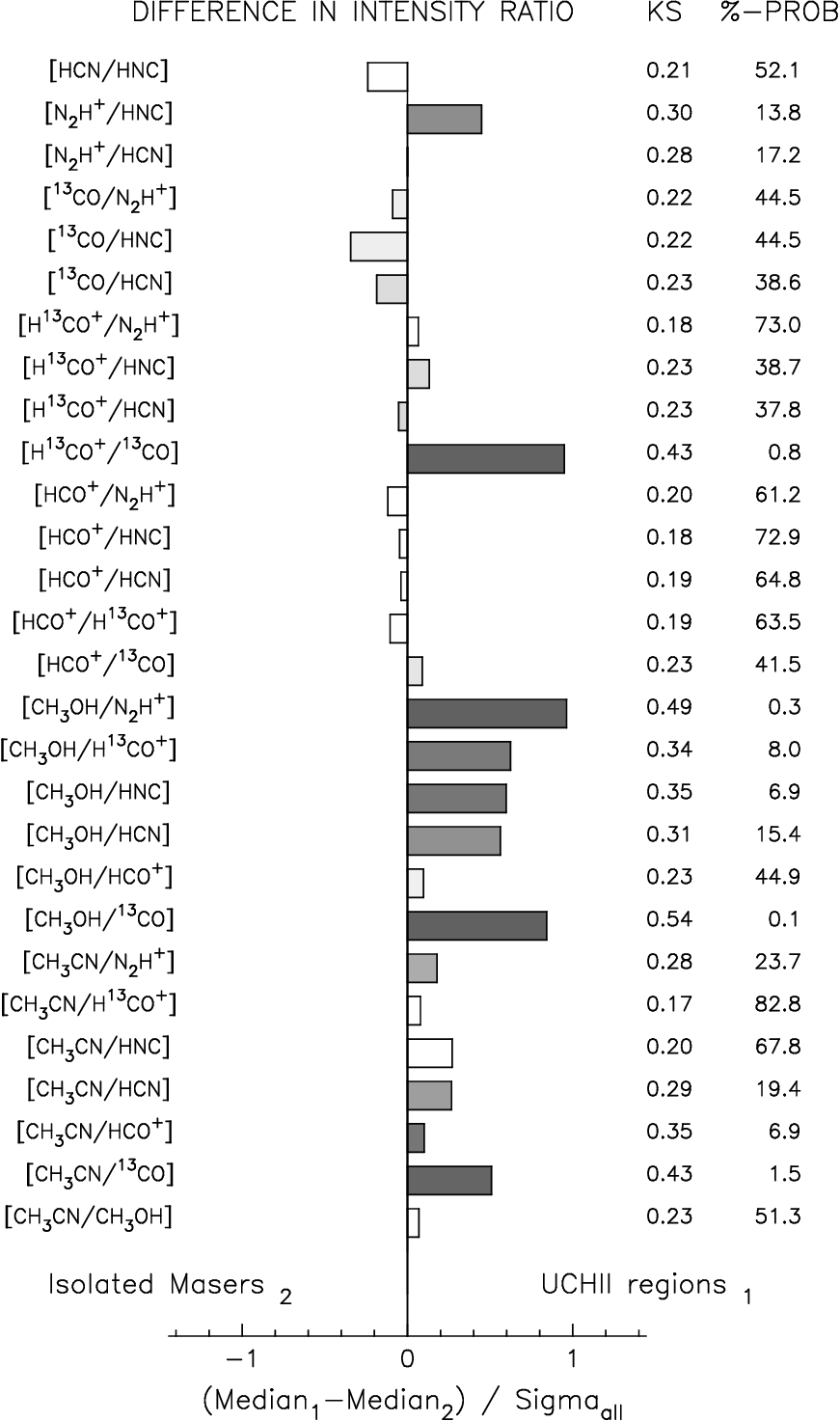}
      \caption{\small~Differences in the median line-line intensity
      ratio for sources with and without an associated
      \uchii~region. For molecules with multiple transitions (e.g.,
      \chthreecn, \chthreeoh) we have summed the intensity of all
      measured lines.}
      \label{fig:diffm_ratios_uchii}
    \end{center}
  \end{minipage}
  \hfill
  \begin{minipage}[t]{8.5cm}
    \begin{center}
      \includegraphics[height=7cm, angle=-90, trim=0 0 -20 0]{figs/wedge_bw.ps}
      \includegraphics[height=7cm, angle=-90, trim=0 0 -30 0]{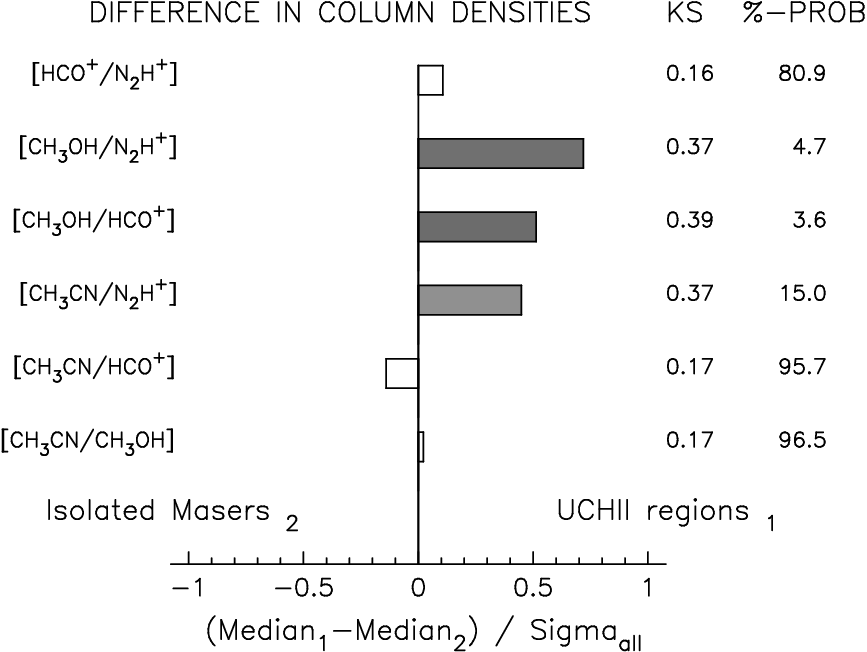}
      \caption{\small~Differences in the median line-line column
      density for sources with and without an associated
      \uchii~region.}
      \label{fig:diffm_column_uchii}
    \end{center}
  \end{minipage}
\end{figure*}
The aim of this molecular survey is to investigate the utility of
molecular lines in the 3-mm band to act as `molecular clocks'. In this
section we consider the differences between sub-samples divided from
the source list based on associations with tracers of potentially
different evolutionary states. 

In Paper 1 we examined the link between hot cores, 6.67\,GHz \chthreeoh~masers
and \uchii~regions. We found that \chthreecn~was brighter and more
commonly detected towards masers associated with \uchii~regions, than
towards `isolated' masers. Assymetries in the \hcop~versus \hthirteencop~line
profiles were interpreted as inward (blue skewed) or outward (red
skewed) gas motions. We also examined the associations with
mid-infrared emission in the Midcourse Space Experiment (MSX)
images and found 11 dark clouds (seen in absorption against the
8\,\micron~Galactic background) and 67 in emission. Many of the
sources also exhibited high velocity line wings, indicative of
outflows. Based on these results we separate the source list into
sub-samples based on the following criteria: 
\begin{itemize}
\item Presence or absence of radio emission from a \uchii~region.
\item \chthreecn~detected versus no \chthreecn~detected.
\item MSX-dark (seen in absorption at 8\,\microns) versus MSX-bright.
\item Presence or absence of high velocity line-wings in \hcop, HNC or
  \thirteenco.
\item Presence or absence of blue-skewed \hcop~line profiles,
  indicating infall.
\end{itemize}
In the following analysis we contrast the sub-samples by looking for
differences in the median properties. A KS-test
is used to assess the similarity of the two distributions and trends
are found based on large differences in the median values, coupled with
a low probability that the sub-populations are drawn from the same
parent distribution.

Initially, we compare the spectral line luminosities calculated from
the integrated line intensity multiplied by the square of the
distance to the source. If the distance is well known then comparisons
of line luminosities will potentially reveal intrinsic differences between
the sub-samples. However, the error in the distance for this sample is
relatively large ($\pm$0.65\,kpc $\approx$ 20 per cent), and increases
drastically towards Galactic longitudes $<$\,5\degree, where the model 
Galactic rotation curve begins to break down. Only the strongest
trends will be seen using such a comparison.

We also compare the distance-independent linewidths.
Linewidths in massive star forming regions are generally
much greater than the thermal width and are indicators of the level
of turbulence present in the gas. Turbulence arises from bulk motions
of gas, is often associated with energetic phenomena, such as
outflows, and is correlated with star formation activity
(e.g. \citealt{Dobbs2005}).

The ratio of two spectral line intensities is unaffected by the
distance and constitutes a much better probe of the physical and
chemical conditions. The dominant errors are due to the relative
accuracy with which the data has been calibrated and the unknown
beam-filling factors. The results presented here are based on
data taken over four years and the calibration error in the data has
been determined to be better than 30 per cent in most cases. The
uncertainty due to the beam filling factors is given by the ratio of
the solid angles over which the two molecules emit,
($\Omega_1/\Omega_2$), and may be large for dense-gas tracers which
subtend angles smaller than the beam. The trends underlying the data
should still be visible given the sample size.

Ideally, in a search for molecular clocks the abundances
should be compared directly. Unfortunately, the small number of
transitions observed in this work make it difficult to constrain the
column densities accurately. Two or more transitions of a single molecule are
needed to solve for the optical depth, excitation temperature and
column density. We have been able to determine the column densities of thermal
\chthreecn~and \chthreeoh~through the rotational diagram  method, and
by assuming an excitation temperature or [$^{12}$C]\,/\,[$^{13}$C] abundance
ratio, we have calculated the column densities of \ntwohp~and
\hcop. Here we compare the beam-averaged column densities of these
molecules directly to find the abundance ratios.

Lastly, we contrast the differences in a suite of measured physical
parameters such as bolometric luminosity, gas-mass,
rotational-temperature and mid-infrared flux density, drawn from this and
other work. 

The important differences are summarised in section~\ref{sec:hmc_2_diff_sum}.


\subsection{Radio-loud versus radio-quiet sources}
\begin{figure}
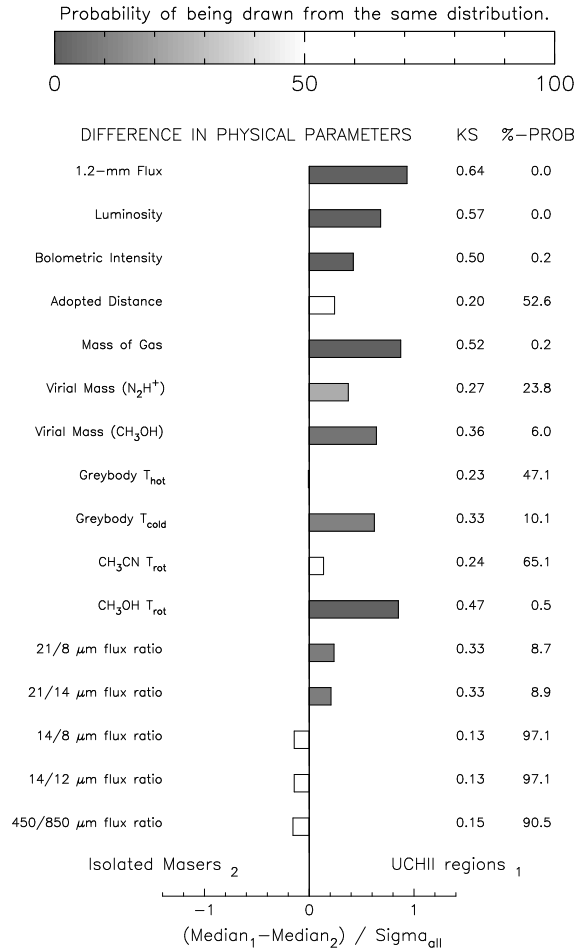

  \begin{minipage}[t]{8.5cm}
    \begin{center}
      \includegraphics[height=7cm, angle=-90, trim=0 0 -20 0]{figs/wedge_bw.ps}
      \includegraphics[height=8cm, angle=-90, trim=0 0 0 -30]{figs/diffm_oth_uchii_bw.ps}
    \end{center}
  \end{minipage}
  \hfill
  \begin{minipage}[t]{8.5cm}
    \begin{center}
      \caption{\small~Differences in other computed and measured parameters,
	for sources with and without an associated \uchii~region.}
      \label{fig:diffm_oth_uchii}
    \end{center}
  \end{minipage}
\end{figure}

Figure~\ref{fig:diffm_intd2_uchii} illustrates the difference in line 
luminosity between sources with and without 8\,GHz radio continuum
emission (from \uchii~regions) in the Mopra beam: the radio-loud and
radio-quiet sub-samples, respectively. As with \chthreecn~and \hcop in
Paper 1, we find that the luminosity of the remaining 
transitions is enhanced towards the radio-loud sub-sample. The median
line luminosity is greater by 0.65-$\sigma$, on average, where
$\sigma$ is the standard deviation of the combined
sample. Thermal \chthreeoh~stands out as having the most significantly
different distribution and the largest difference in
line-luminosity, followed by \hthirteencop, \chthreecn, HNC and
\ntwohp.

We find the median linewidths of all species are broader towards the
radio-loud sub-sample (see Figure~\ref{fig:diffm_lw_uchii}). 
\chthreeoh~exhibits the greatest differences, closely followed by
\ntwohp and \hthirteencop, all of which have statistically different
distributions according to the 
KS-test. \hcop, HCN, HNC and \chthreecn~have broader lines on average,
however the distributions are not significantly different. The
differences in the median linewidth between the radio-loud and
  -quiet subsample range from 0.37\,\kms~to
1.62\,\kms, greater than can be attributed to thermal broadening and
most likely due to the greater turbulent energy in the gas associated
with \uchii~regions. 

The differences found between the linewidths and line-luminosities are
echoed in a comparison of the raw integrated line
intensities. Figure~\ref{fig:diffm_ratios_uchii} shows
the difference in the median values of the 28 intensity ratios between
eight molecules. The  median intensity of \chthreeoh~and, to a
lesser extent, \chthreecn, are clearly enhanced with respect to all
other molecules. Statistically, the most significant difference in
intensity ratio occurs between \chthreeoh~and other molecules,
specifically \ntwohp~and \thirteenco. The [\chthreecn/\thirteenco] and
[\hthirteencop/\thirteenco] ratios also stand out as being
significantly greater towards the radio-loud sub-sample.
\begin{figure*}
  \begin{minipage}[t]{8.5cm}
    \begin{center}
      \includegraphics[height=7cm, angle=-90, trim=-0 0 -10 0]{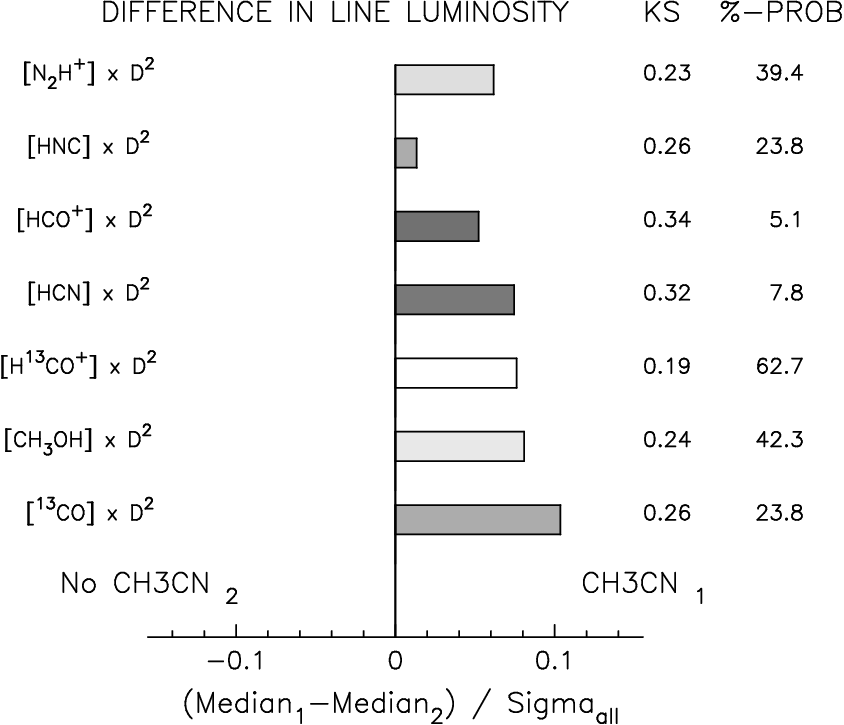}   	
    \end{center}
    \caption[Spectral line luminosities: \chthreecn-bright vs
    \chthreecn-dark.]{\small~Differences in the median line
    luminosity, for sources with and without \chthreecn.}
    \label{fig:diffm_intd2_ch3cn}
  \end{minipage}
  \hfill
  \begin{minipage}[t]{8.5cm}
    \begin{center}
      \includegraphics[height=7cm, angle=-90, trim=-0 0 -5 0]{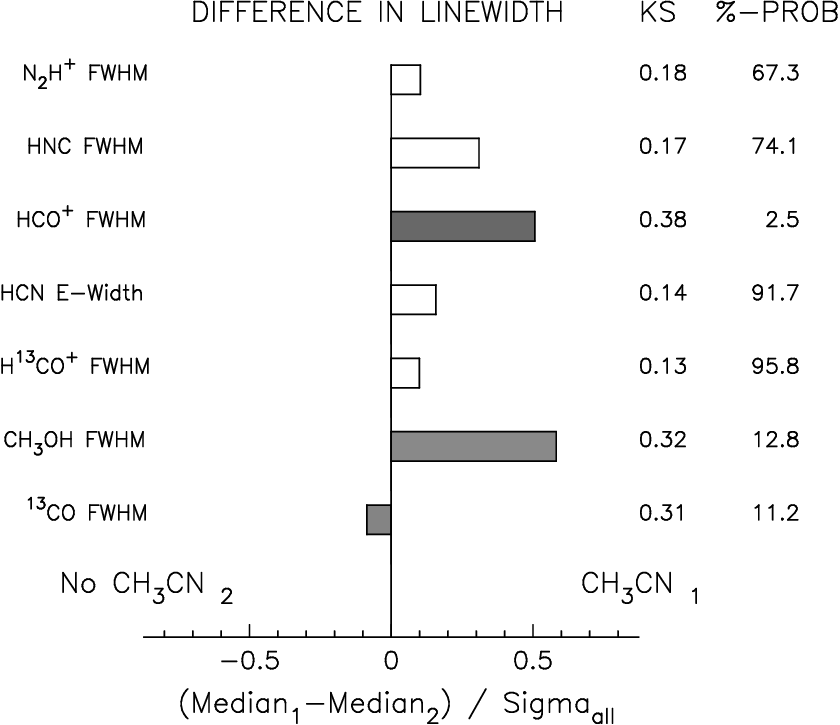}
    \end{center}
    \caption[linewidth: \chthreecn-bright vs \chthreecn-dark.]{\small~Differences in the median linewidth, for sources with and
      without \chthreecn.}
    \begin{center}
    \end{center}
    \label{fig:diffm_lw_ch3cn}
  \end{minipage}
\end{figure*}
\begin{figure*}
  \begin{minipage}[t]{8.5cm}
    \begin{center}
      \includegraphics[height=7cm, angle=-90, trim=0 0 0 0]{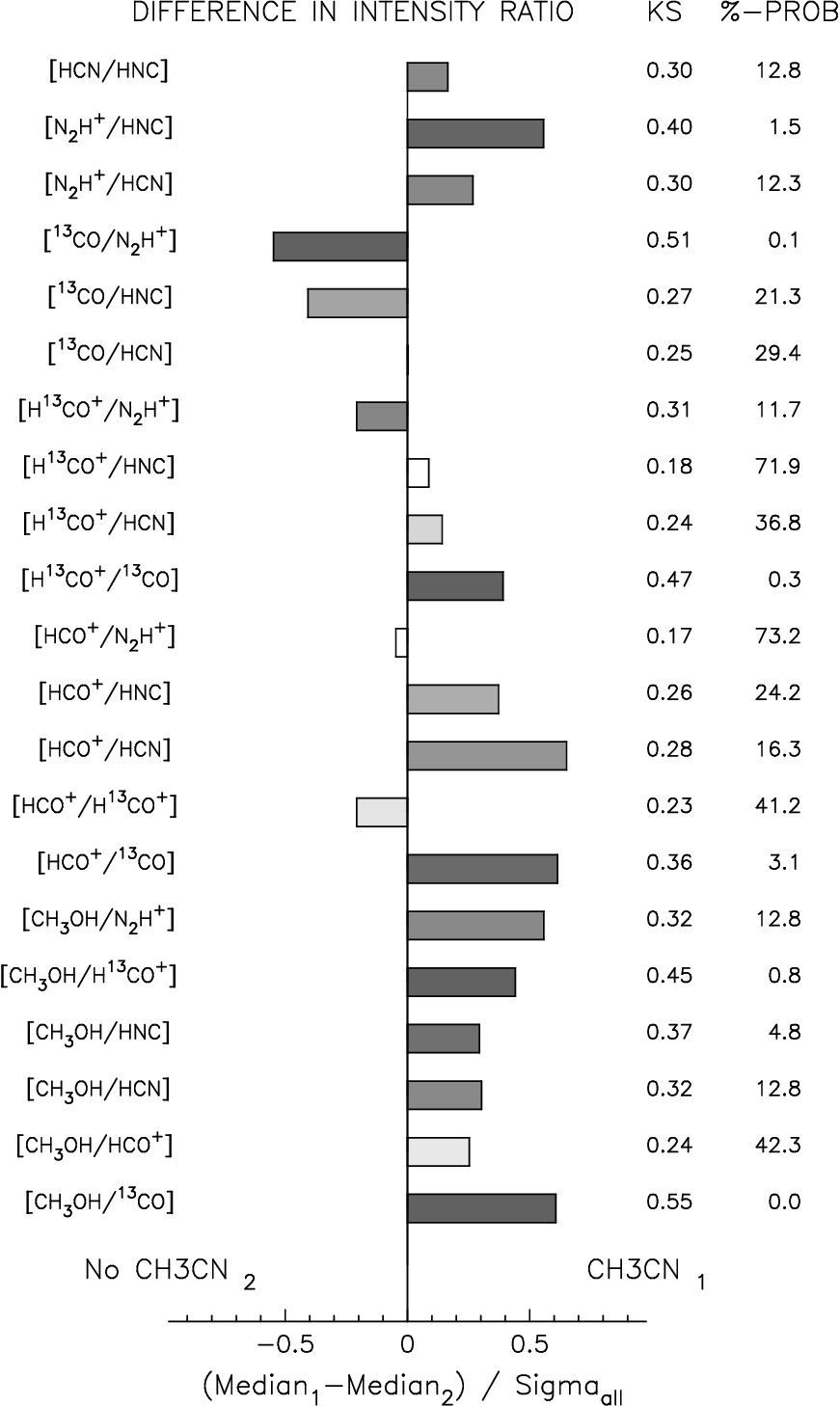}
      \caption{\small~Differences in the median line-line intensity
      ratio for sources with and without \chthreecn. For line with
      multiple transitions we summed the integrated intensity of all
      measured lines.}
      \label{fig:diffm_ratios_ch3cn}
    \end{center}
  \end{minipage}
  \hfill
  \begin{minipage}[t]{8.5cm}
    \begin{center}
      \includegraphics[height=7cm, angle=-90, trim=0 0 -20 0]{figs/wedge_bw.ps}
      \includegraphics[height=7cm, angle=-90, trim=0 0 -30 0]{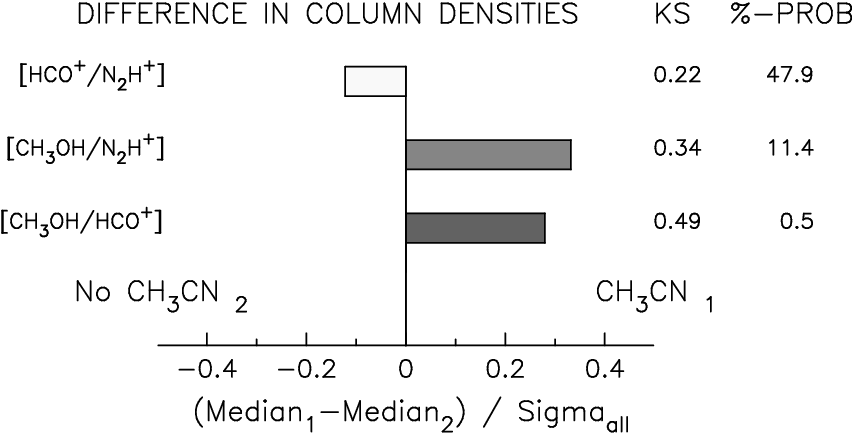}
      \caption[Line intensity ratios: \chthreecn-bright vs
        \chthreecn-dark.]{\small~Differences in the median column
        densities, for sources with and without \chthreecn.} 
      \label{fig:diffm_column_ch3cn}
    \end{center}
  \end{minipage}
\end{figure*}

Figure~\ref{fig:diffm_column_uchii} illustrates a
comparison of the median beam-averaged column densities of four
molecules: \chthreecn, \chthreeoh, \hcop~and \ntwohp. Again, thermal
\chthreeoh~stands out as having greater a column density, and
hence a higher abundance, towards the radio-loud sub-sample.

The median differences between other measured and calculated
properties are shown in Figure~\ref{fig:diffm_oth_uchii}. The median
bolometric luminosity and median mass of the radio-loud sub-sample is
significantly greater than for the isolated masers (see
section~\ref{sec:hmc_lum_mass}). The rotational temperature derived
from \chthreecn~is similar in both groups, however, the
\chthreeoh~rotational temperature is significantly greater in the
radio-bright group (also see Figure~\ref{fig:hist_tdust}).


\subsection{Sources with and without \chthreecn}
\chthreecn~was detected towards 58 sources (70 per cent), however the
detection rate towards the radio-loud sub-sample (19 sources) was
90 per cent. The enhanced luminosities and masses of our \uchii~regions may
bias the results of a comparison between \chthreecn-bright and
\chthreecn-dark groups. We therefore exclude any regions with radio
emission from the comparison. In practise, we find that the major
results are robust to the presence of \uchii~regions.

The differences in median line luminosity between the two groups
presented in Figure~\ref{fig:diffm_intd2_ch3cn} are not as
pronounced as in the previous section, however, all lines are
marginally (0.07-$\sigma$) more luminous towards the \chthreecn-bright
sub-sample. \hcop~and HCN have the most significantly different
distributions. We note that the \hcop~and HCN profiles are affected
by self absorption, with the \chthreecn-bright group containing the
most extreme examples. The significant difference in the
\hcop~luminosity is also not reflected in the optically thin
\hthirteencop~isotopomer, suggesting that an abundance difference is
unlikely. The greatest difference in the median values occurs in
\thirteenco, followed by \chthreeoh~and \hthirteencop.

Linewidths are also biased towards being greater in the sample with
detected \chthreecn~(see
Figure~\ref{fig:diffm_lw_ch3cn}). \hcop~exhibits the most significant
difference, followed by \chthreeoh, although a KS-test
cannot distinguish between the populations with confidence.

When we examine the line intensity ratios in
Figure~\ref{fig:diffm_ratios_ch3cn} a similar picture
emerges to the case of the \uchii~regions. Thermal \chthreeoh~stands out,
as its intensity relative to all other lines is enhanced in the
\chthreecn-bright sub-sample. We also see evidence for enhanced
\ntwohp~emission. In contrast, \thirteenco~emission is relatively weak
towards the \chthreecn-bright sub-sample. These ratios are highly 
suggestive of different environmental conditions in the dense gas,
(traced by \ntwohp~and \chthreeoh) compared to the more extended gas
(traced by \thirteenco), and correlate well with the line
luminosities. 

Figure~\ref{fig:diffm_column_ch3cn} shows the
differences in the median column densities of \chthreeoh, \hcop~and
\ntwohp. Again, the thermal \chthreeoh~column is seen to be
significantly enhanced towards the \chthreecn~bright group, echoing
the result from comparing the raw line ratios. 

\begin{figure*}
  \begin{center}
    \includegraphics[height=6cm, angle=0, trim=0 0 0 0]{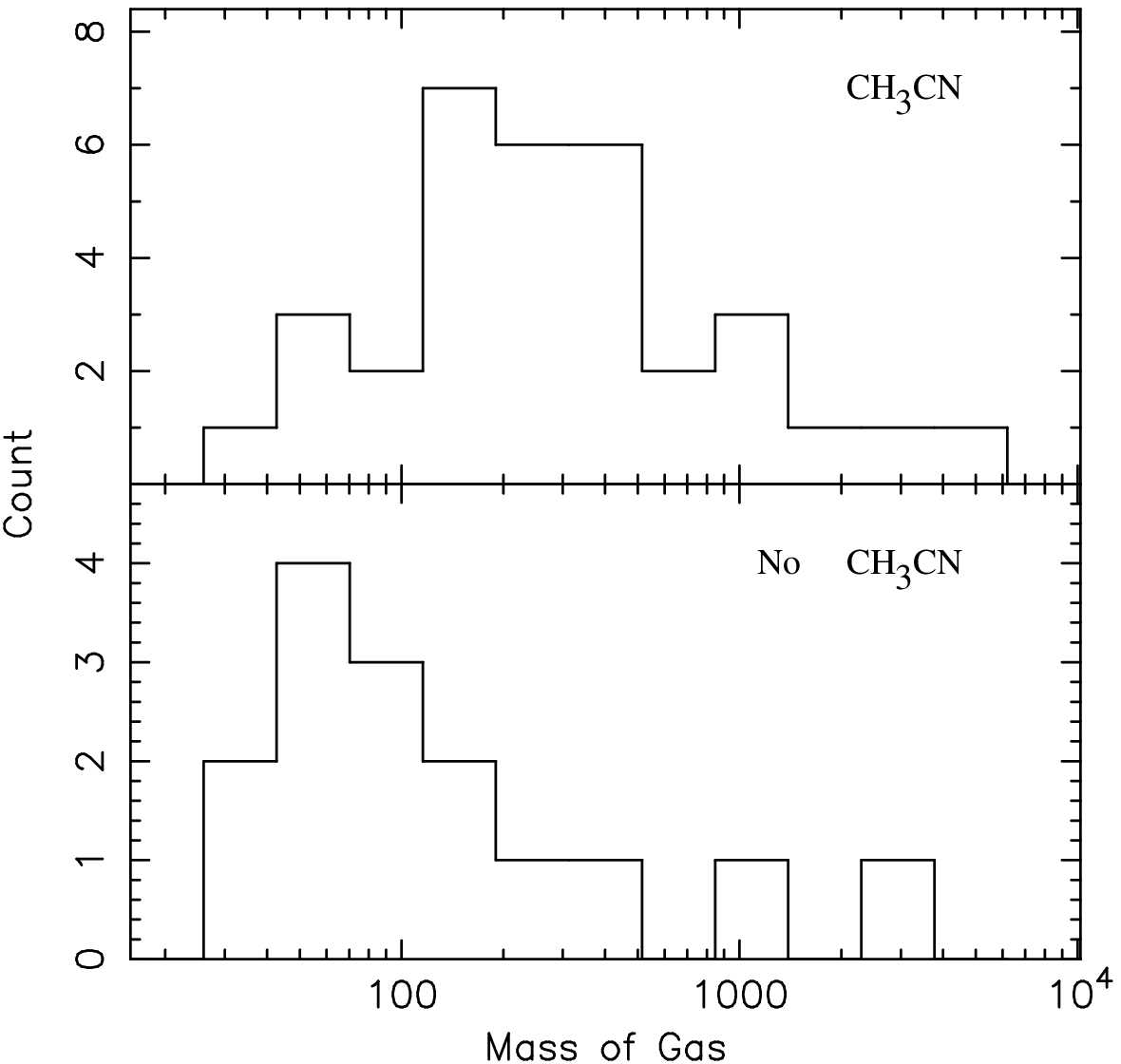}
    \includegraphics[height=6cm, angle=0, trim=-20 0 0 0]{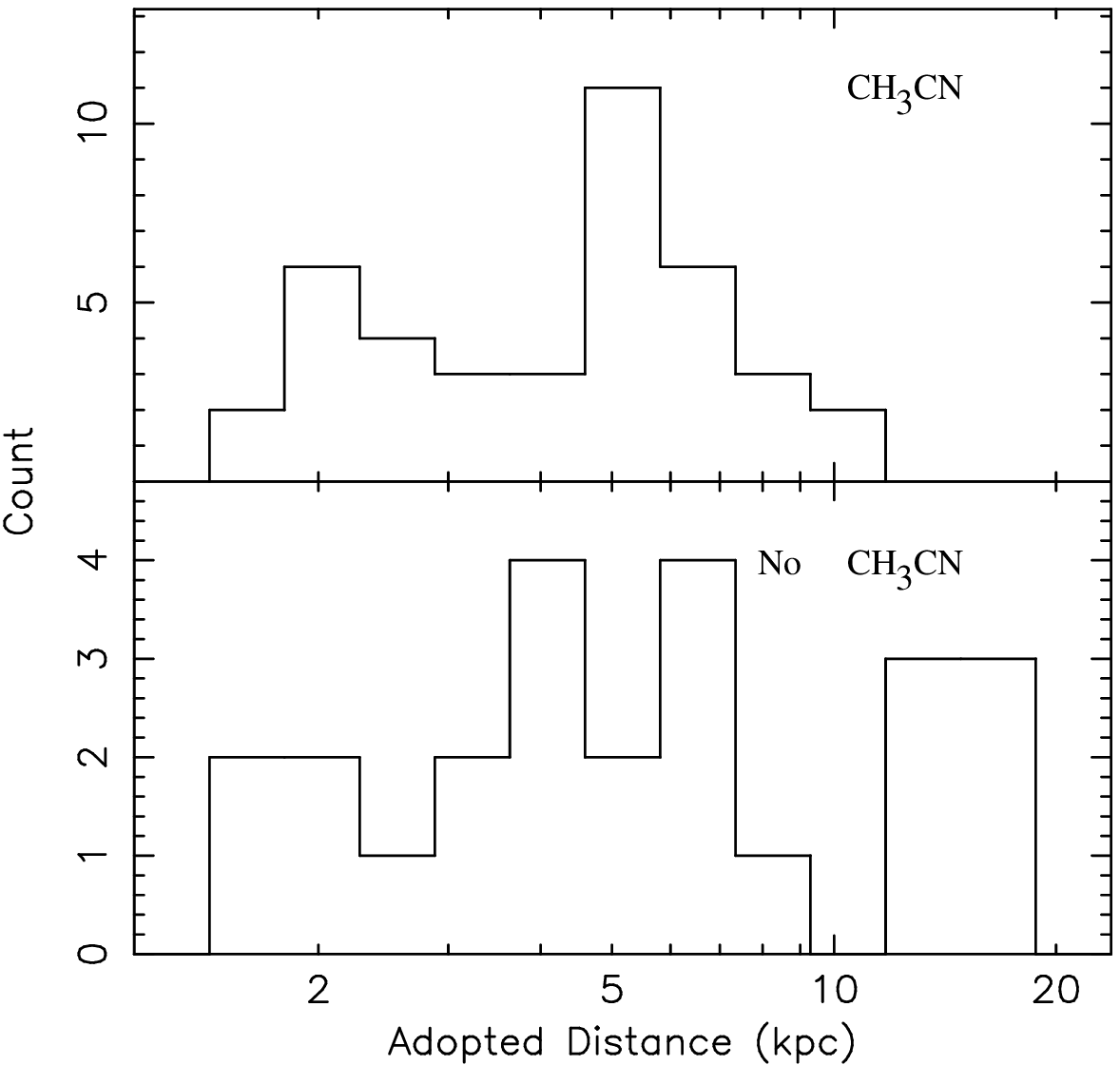}
    \caption[Distributions of gas-masses for \chthreecn~bright and
    dark sources.]{\small~({\it left}) Histograms showing the distributions of
    gas masses derived from 1.2-mm continuum flux density for sources with
    (top) and without (bottom) detected \chthreecn. Sources
    in which \chthreecn~has been detected tend to be more
    massive. ({\it right}) The distribution of distances for sources
    with (top) and without (bottom) detected \chthreecn. A tail of
    more distant objects exists in the \chthreecn-dark sample,
    suggesting that a small fraction of non-detections may be due to
    sensitivity limits.}
    \label{fig:hist_mass_dist_ch3cn}
  \end{center}
\end{figure*}
\begin{figure}
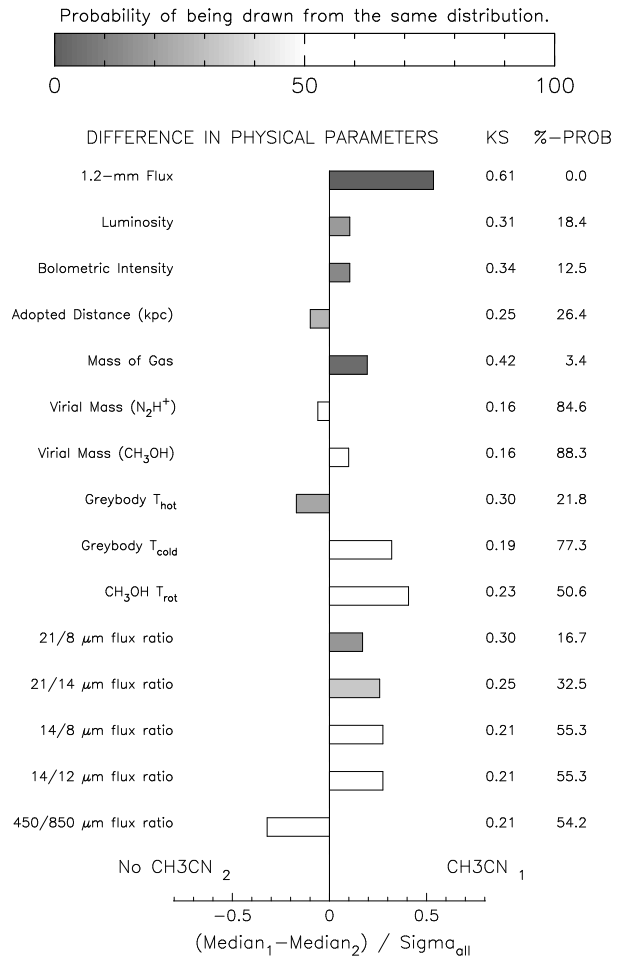

  \begin{minipage}[t]{8.5cm}
    \begin{center}
      \includegraphics[height=7cm, angle=-90, trim=0 0 -20 0]{figs/wedge_bw.ps}
      \includegraphics[height=8cm ,angle=-90, trim=0 0 -30 0]{figs/diffm_oth_ch3cn_bw.ps}
    \end{center}
  \end{minipage}
  \hfill
  \begin{minipage}[t]{8.5cm}
    \begin{center}
      \caption[Other parameters: \chthreecn-bright vs \chthreecn-dark.]{\small~Differences in other computed and measured parameters,
	for sources with and without \chthreecn.}
      \label{fig:diffm_oth_ch3cn}
    \end{center}
  \end{minipage}
\end{figure}

The difference in the median values of other derived parameters are
displayed in Figure~\ref{fig:diffm_oth_ch3cn}. Sources with detected
\chthreecn~have significantly brighter thermal continuum emission at
1.2-mm wavelengths. This translates into a mean mass higher by
$\sim$300\,M$_{\sun}$, omitting \uchii~regions or
$\sim$600\,M$_{\sun}$, including \uchii~regions. Histograms of the
mass distributions for the two groups are plotted in
Figure~\ref{fig:hist_mass_dist_ch3cn} ({\it left}), and show significant
differences.  All other parameters presented in
Figure~\ref{fig:diffm_oth_ch3cn} are similar for both sub-samples. 

The division of the sample into populations with and without detected
\chthreecn~may be a somewhat artificial distinction. While
\chthreecn~emission may be present in more distant sources, it may
fall below our 2-$\sigma$ detection threshold of 80\,mK. Emission over
a small solid angle will suffer from high beam-dilution factors,
further compounding the problem. The distributions of distance in the
two samples, plotted in Figure~\ref{fig:hist_mass_dist_ch3cn} ({\it
  right}), are similar, except for a tail of more distant objects
without \chthreecn. This difference suggests that a small fraction of
non-detections may be attributable to sensitivity limits. 

We also split the sample via other criteria (see the introduction to
this Section), however, the chemical and physical differences
are not as significant. The results of these investigations are
presented as an online supplement.


\subsection{The effect of beam dilution}\label{sec:beam_dilution}
\begin{figure*}
  \centering
  \includegraphics[width=14cm, angle=0, trim=0 0 0 0]{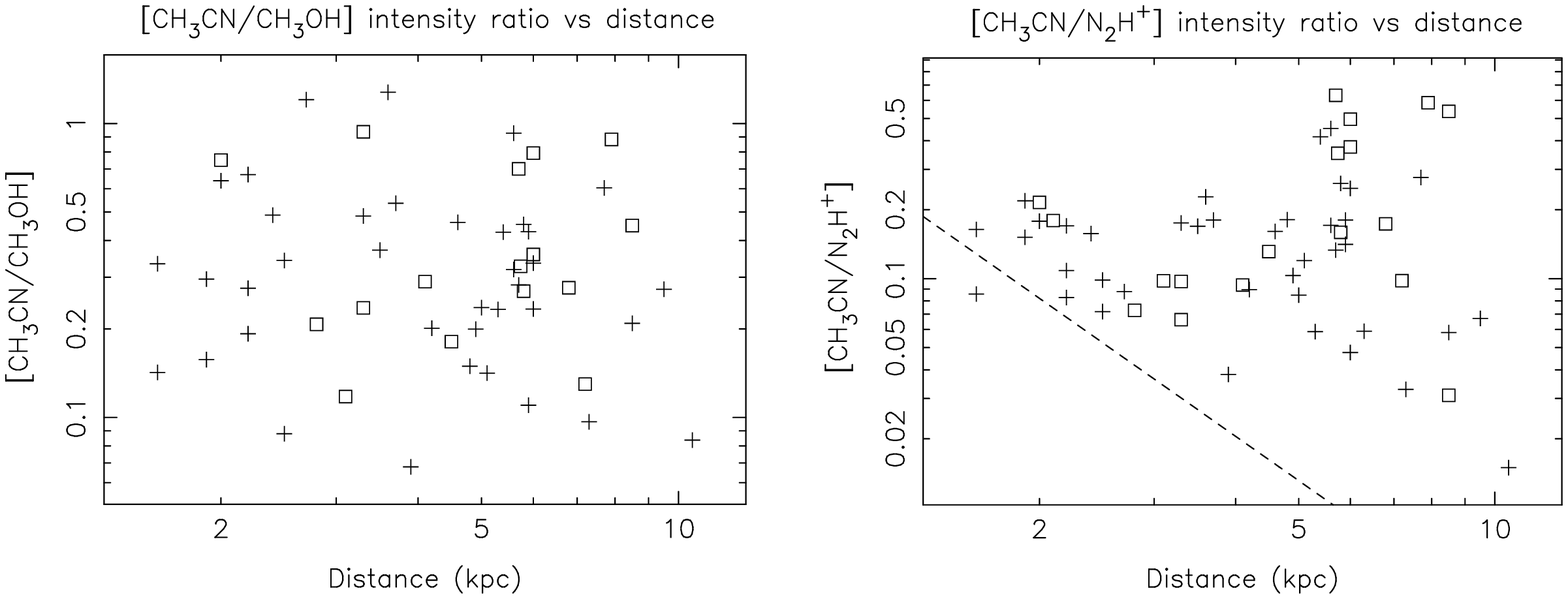}
  \caption[Plots of line-line intensity ratio versus
  distance.]{\small~Plots of line-line intensity ratio versus
  distance. ({\it left}) [\chthreecn/\chthreeoh]~ratio vs
  distance. ({\it right}) [\chthreecn/\ntwohp]~ratio vs
  distance. Crosses mark isolated masers, while squares mark
  \uchii~regions. The dotted line show how the
  [\chthreecn/\ntwohp]~ratio is expected to decrease with distance if
  the \chthreecn~emission has a fixed physical size and the
  \ntwohp~emission always fills the beam.} 
    \label{fig:xy_ratio_dist_bff}
\end{figure*}

As mentioned in Section~\ref{sec:hmc_2_abundance}, the beam filling
factors constitute an unknown source of error when comparing
the intensities or column densities of two molecules. If both emitting
regions subtend angles ($\theta$) smaller than the beam, then we must
correct for the relative volume filling factor, given by
($\theta_1/\theta_2$)$^2$. Assuming the emitting regions 
have similar characteristic sizes in all sources, the relative
beam filling factors will average to $\sim$\,1 over the entire 
sample. However, if one species generally fills the beam while the
other species is generally beam-diluted (i.e., one has constant
angular size while the other doesn't, then we would expect a
1/distance$^2$ dependence in the line intensity
ratios). Figure~\ref{fig:xy_ratio_dist_bff} presents two plots of line 
intensity ratio versus distance. The left panel plots the
\chthreecn/\chthreeoh~ratio versus distance. The emission from both
molecules is expected to derive from regions smaller than the
36\,\arcsec~Mopra beam, even for nearby sources. We see the line-ratios are
not correlated with distance, consistent with a constant ratio between
the angular sizes of the species. The right panel plots the 
\chthreecn/\ntwohp~ratio versus distance. Assuming \ntwohp~emission
closely follows the dust morphology (e.g., \citealt{Caselli2002a, Pirogov2003}),
we have shown in Table~9 that the beam-dilution
factor is generally 
close to 1. In Figure~\ref{fig:xy_ratio_dist_bff} ({\it right}) the
dashed line plots the expected effect of 
\chthreecn~beam dilution as a source is moved further away.
We see the scatter in measured line-ratio increases with
distance making it difficult to interpret the data and, in addition to
beam dilution, chemical and physical differences in the sources affect
the line intensity ratios. Omitting the \uchii~regions (squares) and
considering only isolated masers (crosses), we see a possible trend towards
decreasing \chthreecn/\ntwohp~ratio with increasing distance. This
trend does not approach the 1/distance$^2$ dependence expected, but it
may indicate that some of our comparisons will include beam dilution
effects for \chthreecn.

\subsection{Interpretation and discussion}\label{sec:hmc_2_diff_sum}
It is clear we are seeing significant differences between sources
associated with different tracers. In this section we summarise the
differences between the sub-samples and suggest some possible
interpretations of the results.

\vspace*{2mm}
\noindent{\it \uchii~regions vs isolated masers}\\[1mm]
The clearest distinctions exist between isolated \chthreeoh~masers and
those with \uchii~regions. The result found in Paper 1, where
the \chthreecn~and \hcop~lines were shown to be more luminous and have
greater linewidths in the presence of a \uchii~region, has been
extended to the other observed species. When we examine the raw line
intensity ratios, thermal \chthreeoh~distinguishes itself as being relatively
intense towards the radio loud sources. This is reflected in both the
higher median column density and \chthreeoh~rotational temperature. We
interpret this result as evidence for an enhanced abundance of
\chthreeoh~in relatively warm gas nearby the
\uchii~regions. Significant differences are also found between other
tracers of high and low density gas. The  [\chthreecn/\thirteenco],
[\ntwohp/\thirteenco] and [\hthirteencop/\thirteenco] ratios are
enhanced in the radio-loud sub-sample, possibly indicating different
excitation conditions or abundances in the `core' compared to the
`envelope' gas. Interestingly, when compared to all other lines, the
relative brightness of the hot-core tracer \chthreecn~is {\it not}
significantly greater, yet we detect it towards 95 per cent of
\uchii~regions and only 63 per cent of the isolated masers. 

\begin{figure}
  \begin{center}
    \includegraphics[width=4.4cm, angle=-90, trim=0 -30 -20 0]{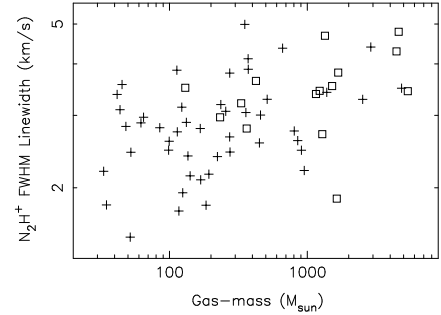}
    \caption[Plot of \ntwohp~linewidth versus
    gas-mass.]{\small~ Plot of \ntwohp~FWHM linewidth versus
    gas-mass. Squares represent \uchii~regions and crosses represent 
    isolated masers.}
    \label{fig:xy_lw_mass}
  \end{center}
\end{figure}

A potential explanation for the median offset in line-luminosity and
linewidths between the radio -quiet and -loud samples may be found in the
significantly greater bolometric luminosity and gas-mass of the
radio-loud sample. Simply put, more 
luminous clumps will likely have a greater flux of IR-photons and
higher gas temperatures. Under these conditions one would expect to
see more luminous spectral lines. Linewidth is commonly used as a
proxy for star formation activity, and hence evolutionary state, as it
is largely dependent on turbulence in massive star forming
regions. However, on the scale of molecular clouds (3-20\,pc) the
linewidth has also been empirically correlated with the cloud mass
\citep{Solomon1987}, possibly due to multiple (sub)-clumps with different
velocities within the beam. It is unclear which effect dominates
in our sample. Figure~\ref{fig:xy_lw_mass} plots the linewidth
\ntwohp~versus the gas-mass. \ntwohp~is chosen as it is optically thin
and has been shown to trace the dense regions of the molecular clouds
(e.g., \citealt{Caselli2002b}). A weak correlation is
observed which may account for some of the linewidth differences
between the groups, however we cannot rule out an evolutionary effect.

That the relative intensity and abundance of \chthreeoh~is greater
towards the radio-loud sources likely reflects real chemical
differences between the sub-samples. This may be interpreted as
evidence for the more advanced evolutionary state of the radio-loud
sample. The \chthreeoh~abundance is predicted to be enhanced by a number
of mechanisms as a massive young stellar object evolves. In the
hot-core phase it is known to be evaporated from the dust grains as
the heating source evolves \citep{Charnley1995}. The abundance of
\chthreeoh~is also predicted to be enhanced in the walls of outflow
cavities, where it is sputtered from the grains by low-velocity shocks
\citep{Hogerheijde2005}.

\vspace*{2mm}
\noindent{\it \chthreecn-bright vs \chthreecn-dark}\\[1mm]
In this and the remaining comparisons we removed the radio-loud sources
from the list, as their higher intrinsic luminosities may bias the
results.
We find sources with detected \chthreecn~are also biased towards
having greater line-luminosities and linewidths, although these
differences are not as pronounced as with the radio-loud versus
radio-quiet division. The most striking result is the significant
tendency for thermal \chthreeoh~to be brighter and have a greater
column-density in the \chthreecn~bright group. This is not unexpected
as the \chthreecn~and \chthreeoh~abundances are predicted to increase
with time in hot cores \citep{Charnley1995}. We also see evidence for
more turbulent and energetic conditions in the 
dense gas (traced by \ntwohp~and \chthreeoh) compared to more extended
and diffuse gas (traced by \thirteenco), implying a more advanced
evolutionary state in the \chthreecn-bright sub-sample. Interestingly,
the \chthreecn-bright sub-sample also has a significantly greater
median mass, which may bias this result. 
\chthreecn~is also not detected in sources with distances
greater than 10\,kpc, suggesting that a small number of sources may
contain \chthreecn~below the sensitivity limit of our survey.

\vspace*{2mm}
\noindent{\it MSX-bright vs MSX-dark}\\[1mm]
All species in the MSX-dark sample have marginally greater
line-luminosities and linewidths compared to their MSX-bright
counterparts, with the \thirteenco~linewidth being especially
notable. This is unexpected, as infrared dark clouds have been
postulated to be the young and quiescent precursors to hot cores and
\uchii~regions \citep{Rathborne2006}. \hcop~and \chthreeoh~tend to be
brighter and \chthreeoh~marginally more 
abundant in the MSX-dark sub-sample, although this is not judged to
be significant. Other parameters are similar in both
sub-samples. Given the small number of sources in the MSX-bright
subsample (9\,--\,11 sources), none of the observed differences are
large enough to draw any definite conclusions from the data.

\vspace*{2mm}
\noindent{\it Line-wings vs no line-wings}\\[1mm]
There are no significant differences between the median line-luminosities
when the sources are divided into sub-samples with and without
high-velocity line-wings. Differences between the \hcop~linewidth
distributions can be attributed to unidentified low-velocity linewings.
\thirteenco~is significantly brighter (relative to other lines) in the
sources without linewings and \chthreecn~is biased towards being
brighter and more abundant in the sub-sample with linewings. No
distinguishing differences are seen in the physical parameters or
IR-colour ratios.

\vspace*{2mm}
\noindent{\it Infall profiles vs no infall profiles}\\[1mm]
The properties of sources with and without \hcop~`infall-profiles' are
close to identical. Little difference is seen between the median
line-luminosities, linewidths or line intensity ratios. Biases in the
\hthirteencop~and \hcop~line ratios can be attributed to selection
effects.

The online material contains additional discusion and figures
pertaining to the previous three paragraphs.


\subsection{Further discussion}
\begin{figure}
  \begin{center}
    \includegraphics[height=6.5cm, angle=-90]{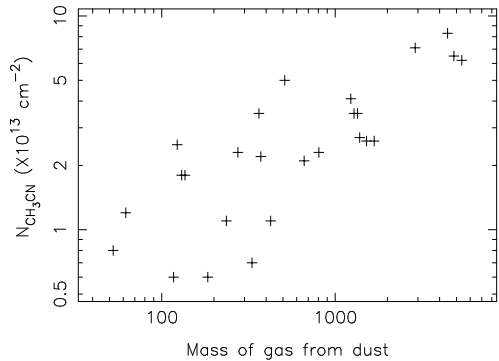}
    \caption[Plot of \chthreecn~column density versus gas-mass.]{\small~Plot of \chthreecn~column density versus gas-mass
    derived from 1.2-mm continuum flux density.}
    \label{fig:xy_Nch3cn_dustmass}
  \end{center}
\end{figure}

Is the mass of the hot core related to the dust mass? 
Figure~\ref{fig:xy_Nch3cn_dustmass} shows a plot of \chthreecn~beam
averaged column density versus mass of gas derived from 1.2-mm
continuum emission. A clear positive correlation is seen between the
two. If we assume the column of \chthreecn~is a proxy for the mass of
the hot core, then it follows that the mass of the hot core increases
with the mass of the extended clump. The relationship is well fit
by M$_{\rm CH_3CN}\,\propto$ M$_{\rm dust}^{0.5}$ and assuming the density
is constant within the gas, this implies that M$_{\rm hot~core}\,\propto$
M$_{\rm dust}^{0.5}$. This is consistent with the
results of \citet{Cesaroni2005} who found that the hot core is an 
continuation of the clump, and there is no break in the approximately
1/r$^2$ density power law. Rather, the hot core is differentiated by
temperature and excitation conditions influenced by the central heating
source.

\section{Summary and conclusions}
In continuation of the `Hot Molecular Cores' survey we used the Mopra
telescope to search for 3-mm transitions of \chthreeoh\,(2\,--\,1),
\ntwohp\,(1\,--\,0), \thirteenco\,(1\,--\,0), HCN\,(1\,--\,0) and
HNC\,(1,--\,0). This is in addition to the previously reported
measurements of \hcop\,(1\,--\,0), \hthirteencop\,(1\,--\,0), and
\chthreecn\,(5\,--\,4) \& (6\,--\,5), in Paper 1. Molecular emission
was detected in all but one source (G10.10$-$0.72), with \ntwohp,
\thirteenco, HNC and HCN detected towards all of the remaining 82
targets. Thermal \chthreeoh~emission was detected in 78 sources (94 per
cent). The following conclusions have been drawn from the data:

\begin{enumerate}
  \item Virial masses estimated from \ntwohp~are comparable to or lower
  than the gas masses of the sample. Values for M$_{\rm gas}$ have
  been calculated from the 1.2-mm dust emission and range from
  $30$\,\msun~to $5\times\,10^3$\,\msun, while  values for M$_{\rm
  virial}$ range from $\sim$\,40 to $2.5\times\,10^3$\,\msun. We note
  that \ntwohp~exhibits particularly narrow linewidths with an average
  of 3.0\,\kms.

  \item H$_2$ volume densities have also been estimated from the
  surface brightness and extent of 1.2-mm emission. The
  mean value is $5\times\,10^4$\,\cmmthree, however, we note that the
  assumptions inherent in the calculation introduce large 
  unknown uncertainties in values for individual sources.

  \item Rotational temperatures derived from \chthreeoh~range from
  3.0\,K to 14.0\,K, with a mean value of 6.67\,K. These anomalously
  low temperatures strongly suggest that the \chthreeoh~is sub-thermally
  excited. Assuming a common excitation temperature, we find the
  abundance ratio of A to E-type \chthreeoh~ranges from 0.5 to 2.8,
  with a mean of 1.5, consistent with previously published values.

  \item \ntwohp~is found to be optically thin towards most of the
  sample. Likewise the \chthreeoh~rotational diagrams
  are consistent with low optical depths in the J\,=\,2\,--\,1 
  transitions. \thirteenco, HCN and HNC all exhibit line profile
  asymmetries and may be significantly optically thick.

  \item The {\it luminosity/gas-mass} relationship for the sample is
  found to follow a rough power law of the form
  L$\propto$\,M$^{0.68}$, meaning that the lower mass clumps are
  over-luminous and the higher mass clumps are under-luminous.
  We find that the gas-mass of the clumps is comparable to or greater
  than the mass of the stellar content (implied by the luminosity). We
  interpret this as an indicator of young ages, as the newborn massive
  stars still retain their natal molecular clouds.

  \item The radio loud sub-sample (\uchii~regions) is on average
  more massive {\it and} more luminous than the radio-quiet sub-sample
  (isolated masers). We argue that the the most likely explanation for
  this disparity is due to the comparatively rapid timescale over
  which the most massive objects evolve into \hii~regions. The first
  significant tracer of star-formation in these objects may be radio-continuum
  emission from a HC\hii~or \uchii~region. It is also
  possible that some of the isolated maser sample contains
  hyper-compact \hii~regions which are too optically thick to detect
  at 8\,GHz. In either case there is a clear age difference between
  the subsamples.

  \item We separated the source list into sub-samples associated with
  different tracers of evolution and compared their median physical
  and chemical properties. Our findings are as follows:

  \begin{itemize}
  \item {\it All} spectral lines are brighter and more luminous towards
    the radio loud sample, extending the result found in
    Paper 1.

  \item The thermal \chthreeoh~brightness and abundance is
  found to be enhanced (compared to other species) in radio-loud
  and \chthreecn-bright sources. We suggest that in single-dish
  observations thermal \chthreeoh~may be used as a crude molecular clock to
  indicate the evolutionary state of a clump.

  \item Given the small sample size, conclusions drawn from a
  comparison of MSX-bright and MSX-dark sources are not significant. It is
  worth noting, however, that linewidths and line-luminosities are
  generally greater in the infrared dark clouds, indicating perhaps
  that when selected using \chthreeoh~masers, dark-clouds
  conceal objects at later stages of evolution.

  \item We divided the source list into sub-samples with and without
  high velocity linewings and \hcop-infall-profiles, but found no
  significant differences in chemical or physical properties.
  \end{itemize}

  \item We find that the beam-averaged \chthreecn~column density is correlated
  with the gas-mass derived from 1.2-mm emission. This suggests that
  the mass of the hot-core is dependent on the mass of the clump in
  which it is embedded.
\end{enumerate}


\section{Acknowledgements}
The authors are greatful to the Australian Research Council and UNSW for
grant support. CRP and SNL were supported by a University of New South
Wales School of Physics Scholarship during the course of this
research. The Mopra radio telescope is part of the Australia Telescope which is
funded by the Commonwealth of Australia for operations as a National
Facility managed by CSIRO. During 2002\,--\,2005 the Mopra telescope
was operated through a collaborative arrangement between the
University of New South Wales and the CSIRO. 

This research has made use of NASA's Astrophysics Data System. 

We would also like thank the anonymous referee for being especially
thorough in seeking out mistakes in the paper. His/her comments were
extremely useful and have helped improve the presentation and interpretation.


\bibliography{paper}

\clearpage
\newpage
\begin{table*}
  \centering
  \begin{minipage}{100mm}
    \caption[Details of observed transitions.]{~Details of observed transitions.}\label{tab:transitions}
    \begin{tabular}{llccccc}
      \hline
      Species       & Transition & Frequency & ${\rm E_{u}/k}$\,$^{\kappa}$ & ${\rm A_{ul}}$ & ${\rm g_{u}}$\\
                    &                         & (GHz)     &    (K)          &(${\rm s^{-1}}$) \\
      \hline  
      \chthreeoh\,$^{\alpha}$            & $2_{(-1,2)}\rarr 1_{(-1,1)}$\,E     & ~~96.739390 & ~~4.642 & 2.495$\times 10^{-6}$ & ~~5\\
                                         & $2_{(0,2)}~~\,\rarr 1_{(0,1)}$\,A+  & ~~96.741420 & ~~6.963 & 3.327$\times 10^{-6}$ & ~~5\\
                                         & $2_{(0,2)}~~\,\rarr 1_{(0,1)}$\,E   & ~~96.744580 &  12.188 & 3.327$\times 10^{-6}$ & ~~5\\
                                         & $2_{(1,1)}~~\,\rarr 1_{(1,0)}$\,E   & ~~96.755510 &  20.108 & 2.496$\times 10^{-6}$ & ~~5\\
      \thirteenco\,$^{\beta}$\rule{0pt}{4mm}& $1\rarr 0$                       &  110.210353 & ~~5.288 & 6.389$\times 10^{-8}$ & ~~3\\
      HNC\,$^{\beta}$\rule{0pt}{4mm}     & $1\rarr 0$                          & ~~90.663593 & ~~4.350 & 2.709$\times 10^{-5}$ & ~~3\\
      HCN\,$^{\gamma}$\rule{0pt}{4mm}    & $1_{1}\rarr 0_{1}$                  & ~~88.630416 & ~~4.253 & 2.444$\times 10^{-5}$ & ~~3\\
                                         & $1_{2}\rarr 0_{1}$                  & ~~88.631847 & --      & --                    & ~~5\\
                                         & $1_{0}\rarr 0_{1}$                  & ~~88.633936 & --      & --                    & ~~1\\
      \ntwohp\,$^{\gamma}$\rule{0pt}{4mm}& $1_{1}\rarr 0_{1}$                  & ~~93.171880 & ~~4.471 & 3.654$\times 10^{-5}$ & ~~9\\
                                         & $1_{2}\rarr 0_{1}$                  & ~~93.173700 & --      & --                    &  15\\
                                         & $1_{0}\rarr 0_{1}$                  & ~~93.176130 & --      & --                    & ~~3\\
      \hline
    \end{tabular}
    \begin{footnotesize}
      $^{\alpha}$~~~${\rm J_{K,K_1}\rightarrow
	J^{\prime}_{K^{\prime},K_1^{\prime}}}$ quantum numbers.

      $^{\beta}$~~~J$\rightarrow$J$^{\prime}$ quantum numbers.

      $^{\gamma}$~~~${\rm J_{F}\rightarrow
	J^{\prime}_{F^{\prime}}}$ quantum numbers. 

      $^{\kappa}$~~\,Energies of the upper levels
      relative to the J$_{\rm K}\,=\,1_{-1}$ level for E-type
      \chthreeoh, and relative to the J$_{\rm K}\,=\,0_{0}$ level for
      A-type \chthreeoh.
    \end{footnotesize}
  \end{minipage}
\end{table*}

\clearpage
\newpage
\begin{table*}
  \centering
    \begin{minipage}{80mm}
      \caption[Noise detection limit on the spectra with no emission.]{~Noise detection limit on the spectra with no emission.}\label{tab:nond_3sig}
      \begin{tabular}{lccccc}
	\hline
	Source \rule[-2mm]{0pt}{1mm} & \multicolumn{5}{c}{T$_{\rm MB}$ 3-$\sigma$ detection limit (mK)} \\
	\cline{2-6}
	\rule[+2mm]{0pt}{1mm}        & \chthreeoh & \thirteenco & \ntwohp & HNC     & HCN     \\	
	\hline
	G0.32$-$0.20                 &   230      & --          & --      & --      & --      \\
	G6.61$-$0.08                 &   104      & --          & --      & --      & --      \\
	G10.10$-$0.72\,$^{\alpha}$    & ~~~69      & 414         & 142     & 127     & 163     \\
	G30.78+0.23                  & ~~~95      & --          & --      & --      & --      \\
	\hline
      \end{tabular}
      \begin{footnotesize}
        $^{\alpha}$~~~Walsh et al. (1997) originally detected a
        6.67\,GHz methanol towards G10.10-0.72. Subsequent
        observations (S. Ellingsen, private communication) have failed
        to detect the maser emission, hence we discard the source from
        ours sample in the remainder of this work.
      \end{footnotesize}
    \end{minipage}
\end{table*}

\clearpage
\newpage
\clearpage
\newpage
\clearpage
\newpage
\clearpage
\newpage
\clearpage
\newpage
\clearpage
\newpage
\clearpage
\newpage

\label{lastpage}
\end{document}